%% file: main.tex
\documentclass[sigconf]{acmart}
\AtBeginDocument{%
  }
\setcopyright{acmlicensed}
\copyrightyear{2018}
\acmYear{2018}
\acmDOI{XXXXXXX.XXXXXXX}
\acmISBN{978-1-4503-XXXX-X/2018/06}

\usepackage{booktabs}
\usepackage{makecell}
\usepackage{multirow}
\usepackage[table,xcdraw]{xcolor}
\usepackage{graphicx}
\usepackage{amsmath}
\usepackage{thmtools}
\usepackage{enumitem}
\usepackage{graphicx}
\usepackage{subcaption}
\usepackage{pifont}
\usepackage{colortbl}
\usepackage{tabularx}
\usepackage[ruled,vlined]{algorithm2e}

\begin{document}

\newcommand{\dk}[1]{{\color{orange}[dk: #1]}}
\newcommand{\cb}[1]{{\color{cyan}[cb: #1]}}
\newcommand{\zr}[1]{{\color{purple}[zr: #1]}}

\title[DIET: Learning to Distill Dataset Continually for Recommender Systems]{DIET: Learning to Distill Dataset Continually for \\ Recommender Systems}

\author{Jiaqing Zhang$^{1}$, Hao Wang$^{1}$$^{*}$, Mingjia Yin$^{1}$, Bo Chen$^{2}$, Qinglin Jia$^{2}$, Rui Zhou$^{1}$, Ruiming Tang$^{2}$, ChaoYi Ma$^{2}$, Enhong Chen$^{1}$}
\affiliation{
    \institution{$^{1}$University of Science and Technology of China, Hefei, China}
    \institution{$^{2}$Kuaishou Technology, Beijing, China}
    \country{}
}

\renewcommand{\shortauthors}{Jiaqing Zhang et al.}

\input{content/abstract}



\maketitle

\input{content/introduction}

\input{content/related_work}
\input{content/preliminary}
\input{content/methodology}
\input{content/experiment}

\input{content/conclusion}

\bibliographystyle{ACM-Reference-Format}
\bibliography{reference}

\appendix
\input{content/appendix}

\end{document}

%% file: content/abstract.tex
\begin{abstract}
Modern deep recommender models are trained under a continual learning paradigm, relying on massive and continuously growing streaming behavioral logs. 
In large-scale platforms, retraining models on full historical data for architecture comparison or iteration is prohibitively expensive, severely slowing down model development. This challenge calls for data-efficient approaches that can faithfully approximate full-data training behavior without repeatedly processing the entire evolving data stream. We formulate this problem as \emph{streaming dataset distillation for recommender systems} and propose \textbf{DIET}, a unified framework that maintains a compact distilled dataset which evolves alongside streaming data while preserving training-critical signals. Unlike existing dataset distillation methods that construct a static distilled set, DIET models distilled data as an evolving training memory and updates it in a stage-wise manner to remain aligned with long-term training dynamics. DIET enables effective continual distillation through principled initialization from influential samples and selective updates guided by influence-aware memory addressing within a bi-level optimization framework. Experiments on large-scale recommendation benchmarks demonstrate that DIET compresses training data to as little as \textbf{1-2\%} of the original size while preserving performance trends consistent with full-data training, reducing model iteration cost by up to \textbf{60$\times$}. Moreover, the distilled datasets produced by DIET generalize well across different model architectures, highlighting streaming dataset distillation as a scalable and reusable data foundation for recommender system development.
\end{abstract}

%% file: content/introduction.tex
\section{INTRODUCTION}

Deep learning recommender models (DLRMs) have achieved substantial performance gains by scaling model capacity and architectural complexity~\citep{Rankmixer-10.1145/3746252.3761507, WuKong-zhang2024wukong, generative-recommender-survey1-10.1145/3701716.3715865, OneRec-zhou2025onerec, generative-recommender-survey2-202512.0203}. 
In modern platforms, recommender models typically operate under a \textbf{continual learning paradigm}, where they are repeatedly updated using \textbf{streaming behavioral logs} generated at the scale of billions of events per day~\citep{Monolith-liu2022monolith}. 
Under this paradigm, adjustments to the model architecture in practice often require retraining on large-scale historical data to obtain reliable performance verification, resulting in steadily increasing iteration costs and slowing down the pace of model design. 
Table~\ref{tab:scaling_cost_perfected} summarizes the growth in training scale and computational requirements of representative DLRMs in recent years. The statistics indicate that as recommender models continue to scale, the computational resources required for full-data training grow rapidly, making large-scale model exploration increasingly costly in practice. 
This trend exacerbates that scalable recommender system development requires not only advances in modeling techniques, but also a reconsideration of how data supports model design.

\input{table/scaling_barrier}

\begin{figure}[t!]
    \centering
    \vspace{0.1\baselineskip}
    \includegraphics[width=1.0\columnwidth]{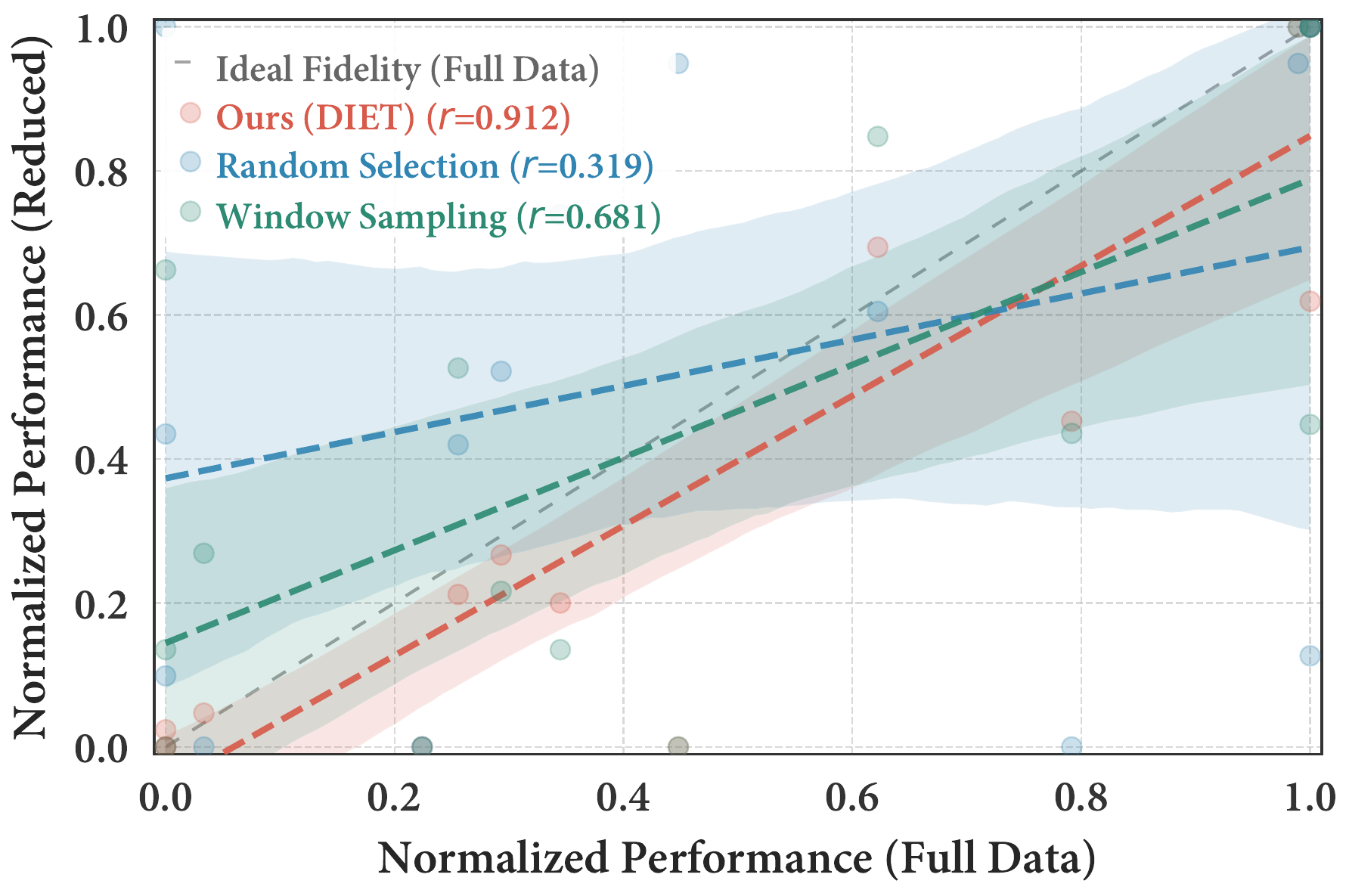}
    \vspace{-2\baselineskip}
    \caption{\textbf{Model performance consistency under data reduction.}
    Correlation between model performance measured on reduced data and on full data.
    Each point corresponds to a model–dataset pair, with the dashed diagonal indicating
    ideal fidelity to full-data training.
    DIET exhibits substantially higher correlation with full-data performance
    than sampling-based baselines, demonstrating superior preservation of comparative
    model behavior.}
    \vspace{-1\baselineskip}
    \label{fig:rankingmap}
\end{figure}

A heuristic approach is to limit the amount of data used for model design, for example by sampling recent behavioral logs or applying coreset selection to construct a smaller training set. 
However, such data reduction strategies often fail to faithfully reflect the training behavior induced by full historical data. 
Recommendation data is typically high-dimensional, sparse, and temporally evolving, and simple data reduction can alter the underlying training data distribution, leading to systematic deviations in the optimization trajectories experienced during training. 
As illustrated in Figure~\ref{fig:rankingmap}, models trained under different sampling strategies exhibit substantial discrepancies compared to those trained on full data, where these inconsistencies reveal an observable consequence of shifted training behavior. 
These observations suggest that the key lies not merely in reducing data volume, but in constructing a compact representation that preserves the training effects of full data. Dataset distillation provides a natural paradigm for achieving this goal.

Unlike subset selection, dataset distillation aims to construct a highly compact memory by learning synthetic samples through optimization, such that training on the distilled dataset can reproduce the training behavior induced by full historical data. 
However, most existing methods are designed for \textbf{static} settings, where the entire training dataset is assumed to be fixed and available in advance, and the distilled dataset is constructed only once~\citep{TD3-zhang2025td3, data-distillation-wang2018dataset, DC-zhao2020dataset}.
In contrast, recommender systems operate under a continual learning paradigm, where streaming behavioral logs are generated continuously and the training data distribution evolve over time. 
Under such conditions, a static distilled dataset is insufficient to capture the evolving training effects required for continual model training. 
Motivated by this gap, we formulate the problem of \emph{\textbf{streaming dataset distillation for recommender systems}}, where distilled data must remain compact while being continuously updated alongside streaming data, so as to consistently approximate the training effects induced by full historical data over time.

Under this setting, streaming dataset distillation introduces several fundamental challenges. 
\textbf{First}, distillation optimization must effectively approximate the training behavior induced by full historical data in recommender systems, where the data space is high-dimensional and informative signals are extremely sparse. 
In such settings, naive or random initialization often fails to provide a meaningful optimization starting point, leading to unstable or ineffective distillation outcomes~\citep{dd-survey1-liu2025evolution,Delt-shen2025delt,du2024diversity}. 
\textbf{Second}, recommender systems operate in a continual learning environment where user interaction data arrive continuously and the underlying data distribution evolves over time. 
Consequently, distilled data must capture the evolving patterns of the original data; otherwise, the distilled samples may gradually drift away from the historical training signals during multi-stage optimization.
\textbf{Finally}, in dataset distillation, synthetic samples are typically treated as learnable parameters and optimized through gradient-based updates. 
Optimizing all distilled samples indiscriminately may introduce redundant or conflicting signals, leading to unstable distillation processes~\citep{difficulty-aligned-dd1-guo2023towards, difficulty-aligned-dd2-lee2024selmatch, difficulty-aligned-dd3-wang2024not}.

To address these challenges, we propose \textbf{DIET}, a unified framework for streaming dataset distillation in recommender systems. 
DIET treats distilled data as a compact training memory that is continuously updated under a continual learning paradigm. 
It adopts a stage-wise distillation strategy that partitions streaming behavioral logs into temporally ordered blocks, enabling progressive distillation while retaining relevant distilled data from previous stages. 
For each stage, DIET initializes new distilled samples from informative real interactions via influence-aware scoring, providing an effective optimization starting point in high-dimensional and sparse settings. 
During optimization, DIET jointly updates newly initialized and retained distilled samples, allowing distilled data to evolve smoothly while preserving previously accumulated training effects. 
Furthermore, DIET incorporates an influence-guided memory addressing mechanism within a bi-level optimization framework to selectively determine which distilled samples should be updated at each stage, ensuring stable and efficient continual updates. Our main contributions are summarized as follows:
\begin{itemize}[leftmargin=1.2em,itemsep=2pt,topsep=2pt]
    \item We are the first to formalize the data-efficiency challenge in recommender systems as \emph{streaming dataset distillation}, where distilled data must evolve over time to consistently approximate the training behavior of full historical interactions.
    
    \item We propose \textbf{DIET}, a continual distillation framework that models synthetic data as evolving training memory, enabling effective initialization in high-dimensional sparse settings, preserving distributional continuity across stages, and supporting stable optimization during continual updates.
    
    \item Extensive experiments show that DIET maintains performance trends consistent with full-data training using only 1--2\% of the data, reduces model iteration cost by up to $60\times$, and produces distilled data that generalize well across model architectures.
\end{itemize}
\vspace{-10pt}

%% file: table/scaling_barrier.tex
\begin{table}[!t]
    \vspace{1.4\baselineskip}
    \centering
    \small
    \caption{
    \textbf{Training Cost Scaling in DLRMs.}
    The upper block reports model statistics from the RankMixer paper~\cite{Rankmixer-10.1145/3746252.3761507}.
    The lower block estimates training cost in GPU hours.
    }
    \vspace{-5pt}
    \label{tab:scaling_cost_perfected}

    \renewcommand{\arraystretch}{1.1}
    \setlength\tabcolsep{3pt} 

    \newcolumntype{C}{>{\centering\arraybackslash}X}

    \begin{tabularx}{\columnwidth}{>{\hsize=1.2\hsize}l | *{3}{C}}
        \specialrule{1.2pt}{1pt}{0pt}
        
        \textbf{Metric} & \textbf{DLRM} & \textbf{Wukong} & \textbf{RankMixer} \\
        \specialrule{0.5pt}{0pt}{0pt}

        \# Parameters ($N$)   & 8.7M   & 122M   & 1B     \\
        Inference Flops ($F$) & 52 G   & 442 G  & 2106 G \\
        MFU ($\eta$)          & 4.51\% & 18.51\%& 44.57\%\\
        
        \specialrule{0.5pt}{1pt}{0pt}
        \multicolumn{4}{>{\columncolor[HTML]{F5F5F5}}l}{
            \rule{0pt}{10pt}\textit{Estimated Training Cost for Scaling Exploration}
        } \\
        \specialrule{0.5pt}{1pt}{0pt}
        
        Sample Scale ($D$) & $10^{11}$ (1d) & $3.0\times10^{12}$ (1m) & $1.5\times10^{13}$ (6m) \\
        Total FLOPs ($C$)  & $5.2\times10^{21}$ & $1.3\times10^{24}$ & $3.2\times10^{25}$ \\
        
        GPU Hours & $\approx 9.7\times10^4$ & $\approx 5.9\times10^6$ & $\approx 6.0\times10^7$ \\
        \specialrule{1.2pt}{1pt}{0pt}
    \end{tabularx}
    
    \vspace{5pt}
    {
    \setlength{\baselineskip}{0.85\baselineskip}
    \textit{Note.}
    Total FLOPs are computed as $C = D \times F$, where $D$ is the number of training samples and $F$ denotes inference FLOPs per sample.
    GPU hours are estimated as
    $T_{\text{hour}} = C / (\eta \cdot P \cdot 3600)$,
    where $\eta$ is the model FLOP utilization (MFU) and $P$ denotes the peak throughput of an RTX~4090 GPU.
    }

\end{table}

%% file: content/related_work.tex
\section{RELATED WORK}

\subsection{TRAINING PARADIGMS IN MODERN RECOMMENDER SYSTEMS}

Modern recommender systems operate on continuously generated, large-scale behavioral logs, where effective modeling requires capturing high-order feature interactions and long-term behavioral dependencies from historical data~\citep{GCRank-ni2025gcrank, 03-wang-00-yin2025feature, 05-wang-02-xu2025multi, 07-wang-05-zhou2025multi, 08-wang-06-wang2025universal, 02-wang-xie2024breaking, 04-wang-01-xie2025breaking}. 
Early studies primarily focused on modeling complex feature interactions through expressive architectural designs~\citep{FM-rendle2010factorization, DeepFM-guo2017deepfm, DCN-wang2017deep, DCNv2-wang2021dcn, AutoInt-song2019autoint, FinalMLP-mao2023finalmlp}. 
Subsequent work shifted attention toward fine-grained interest modeling from behavior sequences~\citep{DIN-zhou2018deep, DIEN-zhou2019deep, BST-chen2019behavior, DSIN-feng2019deep}, while more recent approaches further extend this line of research to model long-term interest evolution under the full historical interaction distribution~\citep{SIM-pi2020search, DV365-lyu2025dv365, TransActv2-xia2025transact}.
In practical systems, achieving such modeling capacity relies on training over continuously accumulated behavioral logs, where model parameters are inherited and updated over time to preserve historical preference signals while adapting to newly observed data~\citep{Monolith-liu2022monolith, Foundation-Expert-Meta-li2025realizing, LUM-yan2025unlocking, OneRec-zhou2025onerec}. 
This continual training paradigm is further reinforced by scaling-law-like phenomena observed in modern recommender models: increasing model capacity, feature dimensionality, and data scale consistently leads to performance improvements~\citep{00-wang-shenp, WuKong-zhang2024wukong, Rankmixer-10.1145/3746252.3761507, FuXi-alpha-ye2025fuxi, LLaTTE-xiong2026llatte}. 
However, under this regime, reliably evaluating new model architectures or training strategies typically requires retraining models on large-scale historical behavioral logs in order to faithfully reflect their training behavior, which in turn results in rapidly escalating iteration costs~\citep{Rankmixer-10.1145/3746252.3761507, HugeCTR-wang2022merlin}. 
These characteristics collectively expose a fundamental challenge in modern recommender system development: how to support continual model iteration without repeatedly retraining on the large-scale historical interaction data.

\subsection{CORESET SELECTION AND DATASET DISTILLATION}

To reduce training cost, prior work explores dataset reduction through \textbf{coreset selection}~\citep{Coreset-har2004coresets} and \textbf{dataset distillation}~\citep{data-distillation-wang2018dataset}. Coreset methods aim to select representative subsets that preserve data diversity or approximate the overall distribution, and have been widely studied in active learning and continual learning~\citep{Coreset-continual-learning-lopez2017gradient, Coreset-active-learning-settles2009active, coreset-continue-learning-aljundi2019gradient, coreset-yang2022dataset}. However, empirical studies show that such methods often provide limited gains over random sampling in deep learning settings~\citep{data-distillation-survey1-sachdeva2023data, data-distillation-survey2-geng2023survey, data-distillation-survey3-liu2025evolution}, as they preserve individual samples but fail to preserve the \emph{training dynamics} induced by the full dataset. Dataset distillation instead synthesizes compact training sets so that models trained on them can approximate the training outcomes on the original dataset~\citep{data-distillation-wang2018dataset}. Representative approaches include bi-level optimization~\citep{data-distillation-wang2018dataset, TD3-zhang2025td3, aat-li2025beyond, ratbptt-feng2023embarassingly}, gradient matching~\citep{DC-zhao2020dataset, DSA-zhao2021dataset, DCctr-wang2023gradient}, and trajectory matching~\citep{MTT1-cazenavette2022dataset, MTT2-du2023minimizing, MTT3-liu2024dataset, MTT4-zhong2025towards}. These methods match training signals such as gradients or parameter states, and thus better preserve training dynamics than selection-based approaches. Existing methods typically assume \textbf{non-streaming settings} with fixed data distributions, and thus fail to capture the training dynamics over time in recommender systems trained on evolving interaction logs.

%% file: content/preliminary.tex
\section{PRELIMINARY}

\paragraph{\textbf{Recommender Data and Continual Training Setup}}

We consider a standard recommender learning setting, where each interaction record is represented as a pair $(\mathbf{x}, \mathbf{y})$.
Here, $\mathbf{x}$ denotes a multi-field feature vector extracted from streaming behavioral logs, and $\mathbf{y} \in \{0,1\}^K$ corresponds to labels of $K$ prediction tasks.
The input features $\mathbf{x}$ are discrete and high-dimensional, and are mapped into a continuous representation space through an embedding layer.
We denote the resulting embedding representation as:
\begin{equation}
\mathbf{e} = \mathrm{Embed}(\mathbf{x}).
\end{equation}
Given a dataset $\mathcal{D} = \{(\mathbf{x}_i, \mathbf{y}_i)\}_{i=1}^{|\mathcal{D}|}$, model parameters $\theta$ are learned by minimizing the empirical multi-task risk:
\begin{equation}
\theta^* =
\arg\min_{\theta}
\frac{1}{|\mathcal{D}|}
\sum_{i=1}^{|\mathcal{D}|}
\sum_{k=1}^{K}
\ell\!\left(
\sigma\!\left(f_\theta(\mathbf{e}_i)_k\right),
y_{i,k}
\right),
\end{equation}
where $f_\theta(\cdot)$ denotes the prediction model operating on embedding representations and outputs a $K$-dimensional prediction vector,
$\ell(\cdot)$ is the binary cross-entropy (BCE) loss,
and $\sigma(\cdot)$ is the sigmoid function applied element-wise.
This formulation provides a unified abstraction for modeling and analyzing the training process of recommender models in the subsequent sections.

In a continual learning paradigm, data arrive sequentially over time.
We partition the full interaction history into a sequence of non-overlapping data blocks
$\mathcal{D}_{1:T} = \{\mathcal{D}_1, \mathcal{D}_2, \dots, \mathcal{D}_T\}$,
each corresponding to a fixed time interval.
Under this paradigm, we distinguish two model roles: a reference model $\phi$ and a candidate model $\theta$. The reference model is continuously evolving, with its parameters inherited across stages and incrementally updated using newly arrived data blocks, while intermediate checkpoints are preserved.
Formally, at stage $t$, the reference model is updated as:
\begin{equation}
\phi_t = \mathrm{Update}(\phi_{t-1}, \mathcal{D}_t),
\quad \phi_0 \sim P(\phi),
\end{equation}
where $\phi_t$ denotes the model state after training on block $\mathcal{D}_t$.
Through this incremental process, the reference model accumulates training signals from the entire interaction history.
In contrast, the candidate model $\theta$ typically corresponds to a candidate architecture or training strategy introduced at a specific stage.
While the reference model has already been trained on historical blocks $\mathcal{D}_{1:t}$, the candidate model only participates in training on subsequent blocks.
This mismatch poses a key challenge: enabling the cantidate model to match the fully evolved reference model without retraining on the full historical data.

\paragraph{\textbf{Streaming Dataset Distillation for Recommender Systems}}

Under the continual learning setting defined above, the goal of streaming dataset distillation is to construct a sequence of compact distilled datasets
$\mathcal{D}^{syn}_{1:T}=\{\mathcal{D}^{syn}_1,\dots,\mathcal{D}^{syn}_T\}$,
such that training on the distilled data substantially reduces computational cost while preserving the training behavior induced by the full historical data.
The distilled dataset is optimized using the reference model and its intermediate checkpoints, while the downstream candidate model is assumed to be unknown at the time of distillation.

Unlike static dataset distillation, streaming dataset distillation must explicitly account for the temporal nature of the data.
At each stage $t$, the distilled dataset $\mathcal{D}^{syn}_t$ is required to reflect the training effects introduced by the current real data block $\mathcal{D}_t$,
while remaining compatible with the training behavior accumulated from earlier stages.
This requirement arises under the continual learning paradigm, where model updates are shaped jointly by historical and newly arriving interactions. Formally, streaming dataset distillation can be characterized as a temporally structured bi-level optimization problem.
At stage $t$, the inner optimization starts from the previous checkpoint $\phi_{t-1}$ and performs $K$ gradient-based updates on the distilled dataset:
\begin{equation}
\phi_t(\mathcal{D}^{syn}_t)
=
\operatorname{Opt}^K\!\left(\phi_{t-1},\ \mathcal{D}^{syn}_t\right),
\end{equation}
where $\operatorname{Opt}^K(\cdot)$ denotes running $K$ steps of the inner-loop optimizer (e.g., SGD) on the distilled dataset.
The outer optimization then updates the distilled dataset to better approximate the training behavior induced by the real data block:
\begin{equation}
\mathcal{D}^{syn*}_t
=
\arg\min_{\mathcal{D}^{syn}_t}
\ \mathcal{L}_{real}\!\left(\phi_t(\mathcal{D}^{syn}_t);\ \mathcal{D}_t\right),
\end{equation}
where $\mathcal{L}_{syn}$ and $\mathcal{L}_{real}$ denote the training losses on distilled and real data, respectively.

%% file: content/methodology.tex
\section{METHODOLOGY}

\begin{figure*}[t!]
    \centering
    \includegraphics[width=0.90\linewidth]{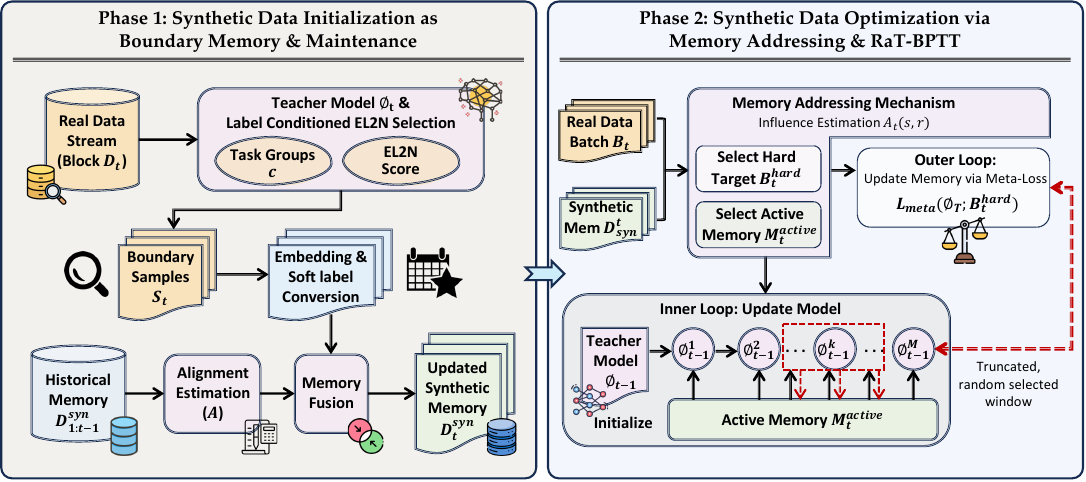}
    \caption{DIET operates in two phases under a continual learning paradigm.
    \emph{Phase~1} (left) constructs a boundary memory by selecting task-conditioned influential samples $\mathcal{S}_t$ from each data block $\mathcal{D}_t$ using reference model checkpoints $\phi_t$ with label-conditioned EL2N scoring.
    The selected samples are converted into embedding--soft label pairs and fused with aligned historical synthetic memory $\mathcal{D}^{syn}_{1:t-1}$ via an alignment estimation module $\mathcal{A}$, yielding the updated synthetic dataset $\mathcal{D}^{syn}_t$.
    \emph{Phase~2} (right) refines the synthetic memory through influence-guided addressing, which selects hard real targets $\mathcal{B}_t^{hard}$ and active synthetic units $\mathcal{M}_t^{active}$.
    The active memory drives inner-loop training of a proxy model initialized from the reference model, while the synthetic data are updated in the outer loop using a meta-objective on $\mathcal{B}_t^{hard}$, accelerated by RaT-BPTT.}
    \label{fig:framework}
\end{figure*}

In this section, we present \textbf{DIET}, a framework for streaming dataset distillation under a continual learning paradigm.
At its core, DIET constructs and continually refines an evolving boundary memory to approximate the training dynamics induced by full historical data.
Figure~\ref{fig:framework} provides an overview of DIET and illustrates how the boundary memory is initialized and continually refined over time.

\subsection{SYNTHETIC DATA INITIALIZATION AS BOUNDARY MEMORY}

Prior studies have shown that training influence is highly concentrated: while most samples contribute redundant gradients, a small subset disproportionately shapes the parameter update trajectory and decision boundary~\citep{EL2N-DBLP:conf/nips/PaulGD21, DBLP:conf/iclr/KaiserSRK25, DBLP:conf/cvpr/ShenS0S25}. 
As a result, constructing distilled data based on global representativeness often fails to preserve the training-critical signals that govern model optimization.
Motivated by this observation, we initialize synthetic data to explicitly capture these training-critical interactions, forming a \textbf{boundary memory}.
Under a continual learning paradigm, this boundary memory must be preserved and updated over time to remain aligned with the evolving training dynamics induced by streaming data. The initialization procedure is summarized in Algorithm~\ref{alg:init}.

\subsubsection{Decision-Boundary Sample Initialization}
\label{sec:4.1.1}

In multi-task recommender systems, interaction data exhibit heterogeneous training influence across both samples and behavioral types.
Different label combinations correspond to semantically distinct user behaviors and often appear with highly imbalanced frequencies.
Directly selecting samples from the entire data block therefore tends to bias synthetic initialization toward dominant behavior modes, while overlooking rare but informative interactions that are critical for shaping the decision boundary.
Therefore, to construct a boundary-aware initialization, we identify \emph{task-conditioned influential samples} within each data block using the reference model checkpoint $\phi_t$.
Specifically, for each possible label configuration $\mathbf{c} \in \{0,1\}^K$, we consider the subset of samples whose label vector equals $\mathbf{c}$:
\[
\mathcal{D}_t^{\mathbf{c}} 
= \{(\mathbf{x}, \mathbf{y}) \in \mathcal{D}_t \mid \mathbf{y} = \mathbf{c}\}.
\]
and rank samples according to their EL2N scores computed with respect to $\phi_t$:
\begin{equation}
\text{EL2N}(\mathbf{x}, \mathbf{y}; \phi_t) =
\left\| \sigma(f_{\phi_t}(\mathrm{Embed}(\mathbf{x}))) - \mathbf{y} \right\|_2.
\end{equation}
Following prior studies that EL2N correlates with gradient influence and sample difficulty~\cite{EL2N-DBLP:conf/nips/PaulGD21}, we use it as a practical proxy for identifying training-influential samples.
To avoid noisy samples associated with extreme scores, we select samples whose EL2N values fall within an upper-middle range for each label group.

The selected samples from all label combinations are aggregated to form the boundary candidate set:
\begin{equation}
\mathcal{S}_t = \bigcup_{\mathbf{c}} \mathcal{S}_t^{\mathbf{c}},
\end{equation}
which ensures balanced coverage across diverse behavioral patterns while preventing dominant labels from overwhelming the initialization.
Instead of storing raw samples, we construct synthetic data in an embedding--soft label form formulated as:
\begin{equation}
\mathbf{e} = \mathrm{Embed}_{\phi_t}(\mathbf{x}),
\quad
\tilde{\mathbf{y}} = f_{\phi_t}(\mathbf{e}),
\end{equation}
where $\mathbf{e}$ denotes the embedding representation and $\tilde{\mathbf{y}}$ is the output logit vector.
The initialized synthetic dataset for stage $t$ is:
\begin{equation}
\mathcal{D}^{syn}_t = \{(\mathbf{e}_i, \tilde{\mathbf{y}}_i)\}_{(\mathbf{x}_i,\mathbf{y}_i)\in \mathcal{S}_t}.
\end{equation}

This initialization strategy captures training-critical samples across heterogeneous behaviors and provides an optimization-aligned starting point for constructing the synthetic boundary memory.

\subsubsection{Continual Synthetic Memory}

Rather than being determined independently within each data block, the decision boundary of a recommender model evolves cumulatively as training proceeds over time.
Initializing synthetic data solely from the current block would therefore capture only a local boundary snapshot and fail to reflect training-influential information accumulated from historical data. To address this issue, we model synthetic data as a \emph{continually evolving boundary memory}.
At stage $t$, we first obtain the task-conditioned training-influential samples $\mathcal{S}_t$ from the current block as described in Sec.~\ref{sec:4.1.1}.
Instead of discarding previously synthesized data, we retain historical synthetic samples that remain aligned with the current decision boundary.
Concretely, for each historical synthetic sample $(\mathbf{e}_i, \tilde{\mathbf{y}}_i) \in \mathcal{D}^{syn}_{1:t-1}$, we estimate its influence weight with respect to the newly selected boundary samples:
\begin{equation}
\alpha_i = \mathcal{A}\!\left((\mathbf{e}_i, \tilde{\mathbf{y}}_i), \mathcal{S}_t\right),
\end{equation}
where $\mathcal{A}(\cdot)$ denotes a influence estimation function whose concrete form will be specified in Sec.~\ref{sec:4.2}.
Historical samples with large $\alpha_i$ form an aligned memory subset $\widetilde{\mathcal{D}}_{syn}^{1:t-1}$.

The synthetic memory for stage $t$ is then constructed by fusing current boundary samples with the aligned historical memory:
\begin{equation}
\mathcal{D}^{syn}_{t} = \mathcal{S}_t \cup \widetilde{\mathcal{D}}_{syn}^{1:t-1}.
\end{equation}

This design enables the boundary memory to evolve smoothly over time by selectively retaining training-influential decision information while incorporating new updates.

\subsection{SYNTHETIC DATA OPTIMIZATION VIA MEMORY ADDRESSING}
\label{sec:4.2}

While boundary-aware initialization and continual retention preserve training-influential information over time, effective distillation further requires selectively updating the boundary memory to remain aligned with evolving training dynamics.
In a continual setting, indiscriminately updating all synthetic samples using all available training samples can introduce redundant or conflicting optimization signals, reducing both efficiency and stability.
The key challenge is therefore to determine which training samples should drive updates and which synthetic memory units should absorb these signals at each stage.
To address this challenge, we introduce an influence-guided memory addressing mechanism that selectively updates the boundary memory using training samples from the current data block, as described next. The optimization procedure is summarized in Algorithm~\ref{alg:opt}.

\subsubsection{Self-Adaptive Optimization Path Discovery}
\label{sec:4.2.1}

\paragraph{Utility and influence estimation.}
For optimization step $t$ with model parameters $w_t$, we define the utility of updating a validation sample $z$ using a subset $S \subset \mathcal{D}$ as:
\begin{equation}
U^{(t)}(S; z) := \ell(\tilde{w}_{t+1}(S), z) - \ell(w_t, z),
\end{equation}
where
\begin{equation}
\tilde{w}_{t+1}(S) = w_t - \eta_t \sum_{x \in S} \nabla_w \ell(w_t, x).
\end{equation}
A smaller $U^{(t)}(S; z)$ indicates a larger loss reduction on $z$.

Applying a first-order Taylor expansion of $\ell(w, z)$ around $w_t$~\citep{TracIN-DBLP:conf/nips/PruthiLKS20,Shapley-DBLP:conf/iclr/WangMS025}, we obtain:
\begin{equation}
\begin{aligned}
U^{(t)}(S; z)
&\approx
\nabla_w \ell(w_t, z)^\top \big(\tilde{w}_{t+1}(S) - w_t\big) \\
&=
-\eta_t \sum_{x \in S}
\left\langle \nabla_w \ell(w_t, z), \nabla_w \ell(w_t, x) \right\rangle .
\end{aligned}
\end{equation}
Under this first-order approximation, the utility is additive over $S$, and the Shapley value of each sample is proportional to its gradient alignment with $z$.
We thus define the influence score as:
\begin{equation}
\mathcal{A}_t(x, z) := \left\langle \nabla_w \ell(w_t, z), \nabla_w \ell(w_t, x) \right\rangle .
\end{equation}

\paragraph{Bidirectional memory addressing.}
Using $\mathcal{A}_t$, we perform bidirectional memory addressing to bridge the gap between real training data and synthetic memory. This mechanism identifies which real samples are most difficult for the current model and which synthetic units are most responsible for representing them.
For each training sample $z \in \mathcal{D}_t$, we define its \emph{alignment deficiency} as the total influence it receives from the current synthetic memory:
\begin{equation}
\text{Deficiency}(z) := \sum_{x \in \mathcal{D}^{syn}_t} \mathcal{A}_t(x, z),
\end{equation}
A smaller value of $\text{Deficiency}(z)$ indicates that the current synthetic memory fails to align with the gradient direction of $z$, making it a hard sample that the model has not yet adequately learned from the distilled data. We select samples with the \textbf{smallest} deficiency values to form the hard target set $\mathcal{B}^{hard}_t$, which serves as the anchor for the outer-loop meta-optimization.
Conversely, we must determine which synthetic units should be updated to minimize the loss on $\mathcal{B}^{hard}_t$. For each synthetic sample $x \in \mathcal{D}^{syn}_t$, we compute its \emph{update responsibility} with respect to the hard targets:
\begin{equation}
\text{Resp}(x) := \sum_{z \in \mathcal{B}^{hard}_t} \mathcal{A}_t(x, z).
\end{equation}
A larger $\text{Resp}(x)$ indicates that the synthetic unit $x$ possesses the highest potential to reduce the meta-loss on the current hard real samples. We select units with the \textbf{largest} responsibility scores as the \emph{active synthetic memory} $\mathcal{M}^{active}_t$.

\subsubsection{Learning Framework}

We optimize the synthetic dataset $\mathcal{D}^{syn}_t$ through a temporal bi-level optimization framework.
The inner loop trains a proxy model on synthetic samples to simulate training dynamics captured by the boundary memory, while the outer loop updates the synthetic data to better align with training samples from the current data block.

\paragraph{Inner-loop Optimization}

At stage $t$, the proxy model parameters $\theta$ are initialized in a stage-aware manner.
For $t=1$, $\theta$ is randomly initialized; for $t>1$, $\theta$ is initialized from the reference model $\phi_{t-1}$ trained on previous stages.
This initialization aligns inner-loop optimization with the training dynamics accumulated over time.
The proxy model is then trained on the active synthetic memory $\mathcal{M}_{t}^{active}$ for $M$ gradient steps:
\begin{equation}
\theta_{m} = \theta_{m-1} - \alpha \nabla_\theta 
\mathcal{L}_{syn}\big(\theta_{m-1}; \mathcal{M}_{t}^{active}\big), 
\quad m=1,\dots,M,
\end{equation}
yielding a simulated optimization trajectory $\{\theta_0, \dots, \theta_M\}$.

\paragraph{Outer-loop Optimization}

Synthetic data quality is evaluated using training samples from the current data block.
Specifically, using the hard target set $\mathcal{B}_t^{hard}$ (Sec.~\ref{sec:4.2.1}), we define the meta objective as:
\begin{equation}
\mathcal{L}_{meta} = 
\mathcal{L}_{real}\big(\theta_M; \mathcal{B}_t^{hard}\big).
\end{equation}
The synthetic dataset $\mathcal{D}^{syn}_t$ is updated by backpropagating through the inner loop, with optimization accelerated using a random truncated backpropagation strategy similar to RaT-BPTT~\citep{ratbptt-feng2023embarassingly}.

\paragraph{Training with Distilled Dataset}
\label{sec:4.2.2}

After completing distillation over historical stages $\mathcal{D}_{1:T}$, the resulting distilled dataset $\mathcal{D}^{syn}_{1:T}$ is used for downstream model training in a warmup-style manner.
Specifically, the dense network parameters of a candidate model are first trained using the distilled data, while sparse embedding parameters are inherited from the reference model at the end of stage $T$.
By combining dense parameters learned from distilled data with the reference model’s embedding representations, we obtain a complete model that preserves historical representation knowledge and enables efficient training without accessing full historical interaction logs. The resulting model can be directly fine-tuned on subsequent data blocks under the continual training paradigm.

%% file: content/experiment.tex
\input{table/dataset}

\section{EXPERIMENT}

\subsection{EXPERIMENTAL SETTINGS}

\input{table/overall}

\subsubsection{Datasets}

We evaluate DIET on three real-world recommendation datasets: \textbf{KuaiRand}\footnote{https://zenodo.org/records/10439422}, \textbf{Tmall}\footnote{https://tianchi.aliyun.com/dataset/dataDetail?dataId=42}, and \textbf{Taobao}\footnote{https://tianchi.aliyun.com/dataset/dataDetail?dataId=649}. These datasets involve multi-field sparse features and multi-task prediction settings. Their overall statistics are summarized in Table~\ref{tab:dataset_stats}. For continual distillation, each dataset is chronologically divided into several historical stages, followed by one subsequent stage and fixed validation and test splits. The historical stages are used for continual dataset distillation and reference model evolution, while the subsequent stage simulates future interactions encountered during continual training. Detailed information refer to Appendix~\ref{app:data_splits}.

\subsubsection{Evaluation Protocols}

We evaluate model performance using two standard metrics for binary prediction tasks: \textbf{Area Under the ROC Curve (AUC)} and \textbf{LogLoss}. 
AUC reflects the ranking quality of predicted probabilities, while LogLoss measures the quality of probability estimation. Under the multi-task setting, we report the average performance across all tasks. For each dataset, models are evaluated on the held-out test split following the continual learning protocol described in Section~5.1.1.

\subsubsection{Baseline and Target Models}

We compare DIET with a set of representative training and data selection strategies under the continual learning setting. \underline{\emph{Full Data}} serves as an upper-bound reference, where models are trained on all historical stages together with the subsequent stage. 
In contrast, \underline{\emph{Cold Start}} trains models only on the subsequent stage, simulating the absence of historical interactions. 
To isolate the benefit of transferring historical representations without accessing historical data, \underline{\emph{Warmup Start}} initializes embedding parameters from the reference model trained on historical stages and continues training on the subsequent stage. We further consider several data selection baselines that compress historical interactions before training. \underline{\emph{Random Selection}} samples data from each historical stage according to the same compression ratio as DIET. \underline{\emph{EL2N Selection}} ranks samples in each historical stage using the EL2N metric and selects those around a chosen percentile (e.g., the 90th percentile) to match the target compression ratio. \underline{\emph{Clustering}} applies K-Means to interaction representations within each historical stage and selects samples closest to cluster centroids to construct a compressed dataset. In all these cases, the selected historical data are combined with the subsequent stage for model training.
To evaluate the generality of DIET across model architectures, we adopt a diverse set of representative recommender models, including WuKong~\citep{WuKong-zhang2024wukong}, DCN~\citep{DCN-wang2017deep}, DCNv2~\citep{DCNv2-wang2021dcn}, FinalNet~\citep{FinalNet-DBLP:conf/sigir/ZhuJCDLDTZ23}, and FinalMLP~\citep{FinalMLP-mao2023finalmlp}. They cover different feature interaction mechanisms and network structures, providing a comprehensive testbed for assessing whether distilled data can consistently support model development across heterogeneous architectures.

\subsubsection{Implementation Details}

All models are implemented within a unified framework based on FuxiCTR~\citep{FuxiCTR-zhu2021open}.
For full-data training, we use a one-epoch setting, which is widely adopted in industrial recommender systems, as models trained on large-scale interaction logs often converge within a single epoch and further passes incur prohibitive cost.
For DIET and all selection-based baselines, models are trained for up to eight epochs on compressed datasets, with the best checkpoint selected based on validation performance. Since the distilled or selected datasets are orders of magnitude smaller, this setting allows sufficient optimization without increasing the overall computational cost.
DIET and all selection-based baselines follow a warmup-style training protocol, as described in Sec.~\ref{sec:4.2.2}, to ensure a fair comparison across methods.
All models are optimized using Adam, with learning rates selected from
$\{1\mathrm{e}{-4}, 3\mathrm{e}{-4}, 1\mathrm{e}{-3}, 3\mathrm{e}{-3}, 1\mathrm{e}{-2}, 2\mathrm{e}{-2}\}$ based on validation performance. For multi-task prediction, all target models use the same embedding dimensionality of 32 and MMoE configuration. More details about distillation process please refer to Appendix~\ref{app:details}.

\subsection{OVERALL PERFORMANCE}

\subsubsection{\textbf{How does DIET compare with selection-based data reduction methods?}}

Table~\ref{tab:final_fixed_alignment} compares DIET with representative selection-based baselines under the same embedding–soft label training protocol.
Despite operating under identical downstream training conditions, DIET consistently outperforms selection-based methods, highlighting the advantage of \emph{dynamic synthesis} over \emph{static selection}.
\textbf{First}, DIET more faithfully approximates full-data training behavior than static subset selection.
On both KuaiRand and Tmall, DIET achieves performance comparable to full-data training using only $\sim$1.5\% of interactions; notably, on KuaiRand, DIET matches the full-data AUC of DCN (0.7878).
In contrast, selection-based methods such as K-Means and EL2N, which rely on fixed subsets of historical samples, exhibit consistent performance gaps, suggesting limited coverage of training-critical signals.
\textbf{Second}, DIET alleviates the limitations of static snapshots in large-scale sparse settings through continual optimization.
On the Taobao dataset, selection-based methods suffer substantial degradation when trained on fixed subsets (e.g., Random Selection: 0.6985 vs. Full Data: 0.7146).
By contrast, DIET continually refines synthetic data via influence-guided optimization, enabling it to retain most of the full-data performance despite extreme data sparsity.

\vspace{-0.5\baselineskip}
\subsubsection{\textbf{How well does distilled data from DIET generalize across model architectures?}}

Experimental results demonstrate that distilled data produced by DIET exhibits strong \textit{cross-architecture generalization}, indicating that it captures training dynamics that are not tied to a specific model structure.
\textbf{First}, DIET enables effective weak-to-strong transfer across architectures.
As shown in Table~\ref{tab:final_fixed_alignment}, although the distilled data is generated using a relatively lightweight DCN reference model, it achieves near-full-data performance when used to train substantially more complex models.
For example, on Tmall, WuKong trained on DCN-distilled data reaches an AUC of 0.7617, closely approaching the full-data upper bound (0.7629) and consistently outperforming selection-based baselines.
This suggests that DIET captures training-relevant patterns that extend beyond the inductive bias of the reference model architecture.
\textbf{Second}, the effectiveness of DIET is largely \emph{reference model-agnostic}.
Figure~\ref{fig:performance_analysis} shows that, on Tmall with a distillation ratio of 0.01, data distilled by different reference models (DCN: 0.7617; FinalNet: 0.7618) yields nearly identical performance when training WuKong, comparable to data distilled by the homologous reference model.
Even at very low ratios (e.g., $10^{-4}$), the performance differences across reference model architectures remain minimal, indicating that DIET optimizes synthetic data toward architecture-invariant training dynamics rather than specific signals.

\vspace{-0.8\baselineskip}
\subsection{FURTHER ANALYSIS}

\subsubsection{\textbf{How do reference model capacity and distillation ratio affect the distilled data?}}

\begin{figure}[t!]
    \centering
    \includegraphics[width=0.45\textwidth]{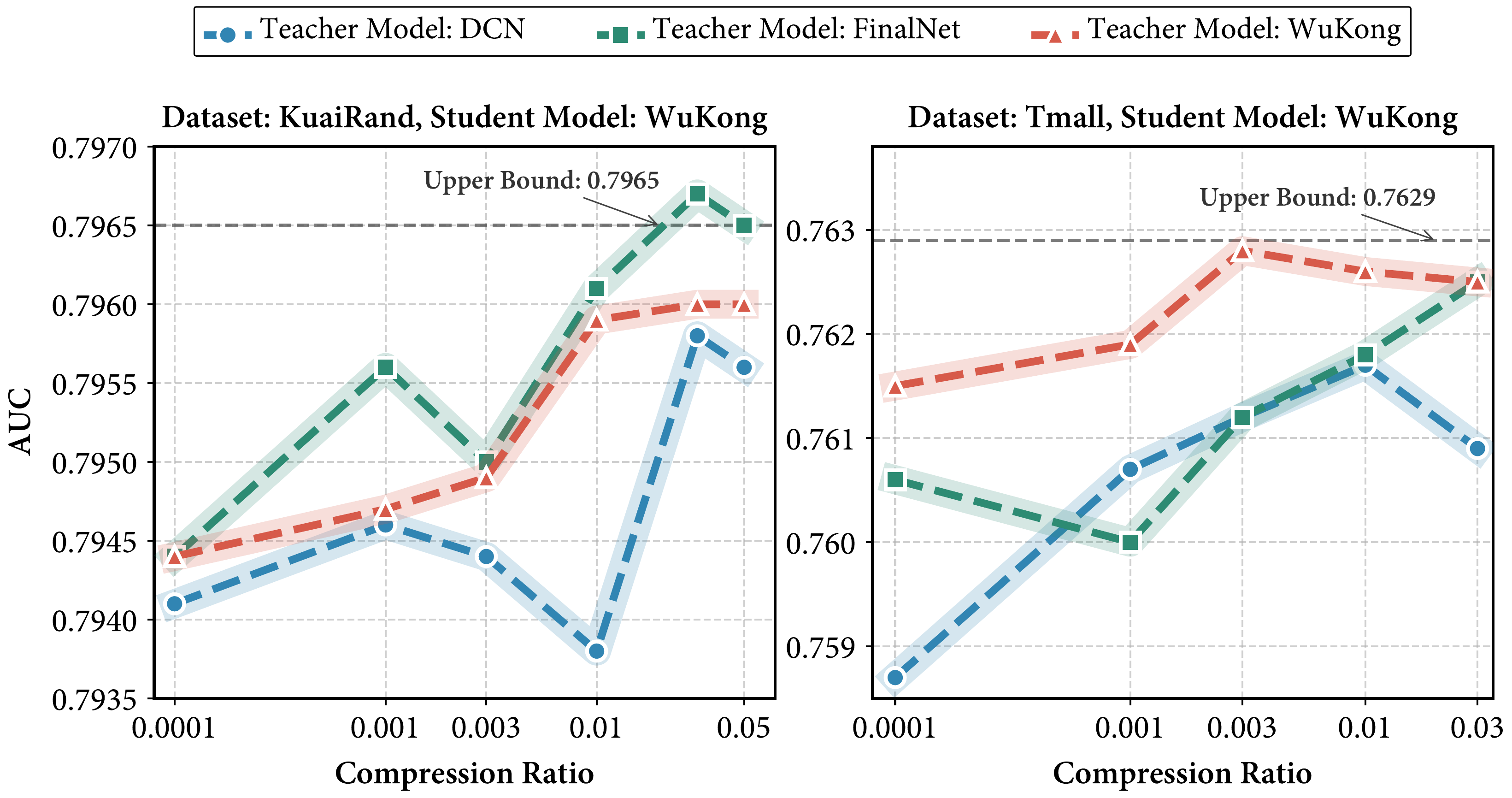}
    \caption{Performance comparison across compression ratios with \textbf{WuKong} as the candidate model. Subplots (a) and (b) show results on the KuaiRand and Tmall datasets. The dotted lines represent the full-data upper bound.}
    \label{fig:performance_analysis}
\end{figure}

Figure~\ref{fig:performance_analysis} examines how distillation ratio and reference model capacity.
\textbf{First}, the candidate model exhibits clear \textit{diminishing marginal returns} as the distillation ratio increases.
On the Tmall dataset, using only $0.3\%$ distilled data already yields an AUC of $0.7628$, recovering over $99.9\%$ of the empirical full-data upper bound ($0.7629$).
This indicates that DIET effectively removes redundancy from massive interaction logs and concentrates training-critical signals into a compact synthetic memory.
\textbf{Second}, \textit{reference model capacity} plays a more decisive role than architectural homogeneity in determining distilled data quality.
Although WuKong and FinalNet differ substantially in architecture, data distilled by the higher-capacity FinalNet consistently outperforms that distilled by DCN, and even exceeds data distilled by the homogeneous WuKong reference model at moderate ratios (e.g., $>0.01$ on KuaiRand).
This suggests that a stronger reference model provides a more accurate approximation of the underlying training dynamics, yielding distilled data with robust \textit{cross-architecture transferability}.
\textbf{Finally}, on the highly dense KuaiRand dataset (average $>8{,}000$ interactions per user), training with $3\%$ data distilled by FinalNet slightly exceeds the full-data reference performance (AUC $0.7967$ vs. $0.7965$).
We attribute this behavior to an implicit \textit{denoising effect}: the strong reference model suppresses spurious or noisy interactions during distillation, resulting in a synthetic dataset that better captures the dominant training signals than the full raw historical data stream.

\input{table/ablation}
\vspace{-5pt}

\subsubsection{\textbf{How do different components of DIET contribute to its performance?}}

The ablation results in Table~\ref{tab:ablation_optimized} indicate that DIET’s performance gains arise from the complementary effects of continual memory fusion and influence-guided addressing.
\textbf{First, Continual Memory Fusion (MF) is essential for handling evolving long-tail distributions.}
Removing MF—i.e., initializing synthetic data solely from the current block without retaining historical memory—consistently degrades performance across all datasets. The effect is most pronounced on the large-scale Taobao dataset, where AUC drops from 0.7142 to 0.7108, suggesting that relying on local data snapshots alone fails to preserve training-influential structure accumulated over time.
\textbf{Second, Influence-guided Bidirectional Addressing (TS \& MS) is crucial for precise and stable optimization.}
Ablating either Target Selection (TS) or Memory Selection (MS) leads to inferior results. In particular, removing MS causes the largest drop on KuaiRand (0.7959 $\to$ 0.7930), indicating that indiscriminate updates to synthetic memory reduce optimization stability. These results highlight the importance of jointly selecting informative real samples and selectively updating the corresponding synthetic memory units in a continual distillation setting.

\input{table/complexity}
\subsubsection{\textbf{How efficient is DIET in terms of time and memory complexity?}}

We elucidate the theoretical computational footprint of DIET in terms of a single outer-loop optimization step. Note that Phase 1 is a one-pass initialization with linear complexity relative to data size and is thus omitted. The empirical runtime is reported in Table~\ref{tab:empirical_complexity}.
Notably, the distillation overhead is incurred only once and can be amortized over multiple downstream model training runs, resulting in a negligible marginal cost in practice.

\paragraph{Memory Complexity.}
The memory consumption $\mathcal{C}_{mem}$ is dominated by the model parameters, the synthetic memory, and the cache required for backpropagation:
\begin{equation}
    \mathcal{C}_{mem} \approx 
    \underbrace{|\theta|}_{\text{Model}} + 
    \underbrace{N_{syn} \cdot (d + L)}_{\text{Synthetic Memory}} + 
    \underbrace{B_{real} \cdot d}_{\text{Real Data}} + 
    \underbrace{K_{rat} \cdot |\theta|}_{\text{RaT-BPTT Cache}}
\end{equation}
where $d$ denotes the feature dimension. By optimizing embeddings directly, the storage cost $N_{syn} \cdot (d+L)$ remains independent of raw data sparsity. Furthermore, the use of RaT-BPTT limits the differentiation cache to $K_{rat}$ steps, where $K_{rat} \ll M$.

\paragraph{Time Complexity.}
The runtime per outer iteration $\mathcal{C}_{time}$ comprises influence-guided addressing, inner-loop simulation, and meta-gradient computation:
\begin{equation}
    \mathcal{C}_{time} \approx 
    \underbrace{(B_{real} + N_{syn}) \cdot |\theta|}_{\text{Memory Addressing}} + 
    \underbrace{M \cdot B_{active} \cdot |\theta|}_{\text{Inner-Loop Training}} + 
    \underbrace{K_{rat} \cdot B_{active} \cdot |\theta|}_{\text{Meta-Gradient}}
\end{equation}
A key efficiency design in DIET is the \textbf{Memory Addressing} mechanism. By selecting only the most relevant synthetic units, the inner-loop cost scales with the small active subset size $B_{active}$ rather than the full memory size $N_{syn}$ (i.e., $M \cdot B_{active} \ll M \cdot N_{syn}$), which significantly reduces the computational overhead per iteration.

%% file: table/dataset.tex
\begin{table}[t]
    \centering
    \small
    \renewcommand{\arraystretch}{1.0}
    \setlength{\tabcolsep}{1pt}
    \caption{\textbf{Statistics of the Datasets.}
    Overview of dataset scale, feature fields, and continual training setup.
    Each dataset is chronologically split into historical blocks for distillation
    and a subsequent block for simulating future interactions.}
    \label{tab:dataset_stats}

    \vspace{-5pt}

    \begin{tabular*}{0.9\columnwidth}{@{\extracolsep{\fill}}cc|ccc|cc@{}}
        \specialrule{1.2pt}{1pt}{0pt} 
        \multicolumn{2}{c|}{\textbf{Features}} &
        \multicolumn{3}{c|}{\textbf{Scale}} &
        \multicolumn{2}{c}{\textbf{Training}} \\
        \cmidrule(lr){1-2} \cmidrule(lr){3-5} \cmidrule(lr){6-7}
        Fields & Tasks & Users & Items & Inters. & Span & Blocks \\
        \specialrule{0.5pt}{1pt}{1pt} 

        \multicolumn{7}{>{\columncolor[HTML]{F5F5F5}}l}{
            \rule{0pt}{6pt}\textit{Tmall Dataset}
        } \\
        \specialrule{0.5pt}{1pt}{1pt} 
        8 & 3 & 379,603 & 675,597 & 19,573,214 & 125d & 2+1 \\
        \specialrule{0.5pt}{1pt}{1pt} 

        \multicolumn{7}{>{\columncolor[HTML]{F5F5F5}}l}{
            \rule{0pt}{6pt}\textit{Taobao Dataset}
        } \\
        \specialrule{0.5pt}{1pt}{1pt} 
        4 & 3 & 987,991 & 4,161,138 & 100,095,231 & 9d & 3+1 \\
        \specialrule{0.5pt}{1pt}{1pt} 
        
        \multicolumn{7}{>{\columncolor[HTML]{F5F5F5}}l}{
            \rule{0pt}{6pt}\textit{KuaiRand Dataset}
        } \\
        \specialrule{0.5pt}{1pt}{1pt} 
        17 & 3 & 1,000 & 3,403,538 & 8,799,733 & 23d & 3+1 \\

        \specialrule{1.2pt}{1pt}{0pt} 
    \end{tabular*}
\end{table}

%% file: table/overall.tex
\begin{table*}[t]
    \centering
    \vspace{-0.5\baselineskip}
    \small
    
    \definecolor{VictimBrown}{HTML}{EFEBE7}
    \definecolor{OursBlue}{HTML}{EEF3FA}
    \newcommand{\vb}{\cellcolor{VictimBrown}}
    \newcommand{\ob}{\cellcolor{OursBlue}}
    \newcommand{\best}[1]{\textbf{#1}}

    \caption{\textbf{Overall performance of DIET} compared with heuristic and selection-based data reduction methods, and its generalization across target models with different architectures. Distilled datasets are generated using DCN as the teacher model and applied to train DCN, DCNv2, FinalNet, FinalMLP, and WuKong. The compression ratio of distilled data is reported next to each dataset name. Best AUC results are in \textbf{bold}, second-best underlined.}

    \vspace{-0.5\baselineskip}
    
    \label{tab:final_fixed_alignment}

    \renewcommand{\arraystretch}{1.1} 
    \setlength\tabcolsep{1.5pt} 

    \begin{tabularx}{0.93\textwidth}{l | *{10}{>{\centering\arraybackslash}X}}
    \specialrule{1.2pt}{0pt}{1pt}
    
    & \multicolumn{2}{c}{\textbf{DCN}} & \multicolumn{2}{c}{\textbf{DCNv2}} & \multicolumn{2}{c}{\textbf{FinalNet}} & \multicolumn{2}{c}{\textbf{FinalMLP}} & \multicolumn{2}{c}{\textbf{WuKong}} \\
    \cmidrule(lr){2-3} \cmidrule(lr){4-5} \cmidrule(lr){6-7} \cmidrule(lr){8-9} \cmidrule(lr){10-11}
    
    Method & AUC $\uparrow$ & LogLoss $\downarrow$ & AUC $\uparrow$ & LogLoss $\downarrow$ & AUC $\uparrow$ & LogLoss $\downarrow$ & AUC $\uparrow$ & LogLoss $\downarrow$ & AUC $\uparrow$ & LogLoss $\downarrow$ \\
    \specialrule{0.5pt}{1pt}{0pt}

    \multicolumn{11}{c}{\textit{KuaiRand Dataset} (1.50\%)} \\ 
    \specialrule{0.5pt}{0pt}{0pt}
    \best{Full Data}   & \vb \textbf{0.7878} & \vb 0.3837 & \vb \underline{0.7931} & \vb 0.3834 & \vb \textbf{0.7898} & \vb 0.3815 & \vb \textbf{0.7875} & \vb 0.3819 & \vb \textbf{0.7965} & \vb 0.3804 \\
    Cold Start         & 0.7701 & 0.3915 & 0.7800 & 0.3857 & 0.7745 & 0.3851 & 0.7655 & 0.3878 & 0.7826 & 0.3868 \\
    Warmup Start       & 0.7822 & 0.3869 & 0.7888 & 0.3841 & 0.7878 & 0.3859 & 0.7847 & 0.3847 & 0.7925 & 0.3829 \\
    Random Selection   & 0.7848 & 0.3866 & 0.7897 & 0.3841 & 0.7882 & 0.3855 & 0.7856 & 0.3862 & 0.7929 & 0.3852 \\
    K-Means Clustering & 0.7846 & 0.3870 & 0.7919 & 0.3827 & 0.7877 & 0.3859 & 0.7853 & 0.3837 & 0.7917 & 0.3817 \\
    EL2N Selection     & 0.7854 & 0.3853 & 0.7911 & 0.3814 & 0.7889 & 0.3820 & 0.7867 & 0.3846 & 0.7925 & 0.3817 \\
    \best{Ours (DIET)} & \ob \textbf{0.7878} & \ob 0.3831 & \ob \textbf{0.7933} & \ob 0.3808 & \ob \underline{0.7892} & \ob 0.3841 & \ob \underline{0.7874} & \ob 0.3835 & \ob \underline{0.7959} & \ob 0.3817 \\
    
    \specialrule{0.5pt}{1pt}{0pt}

    \multicolumn{11}{c}{\textit{Tmall Dataset} (1.39\%)} \\ 
    \specialrule{0.5pt}{0pt}{0pt}
    \best{Full Data}   & \vb \underline{0.7571} & \vb 0.2410 & \vb \textbf{0.7591} & \vb 0.2405 & \vb \textbf{0.7584} & \vb 0.2406 & \vb \textbf{0.7588} & \vb 0.2405 & \vb \textbf{0.7629} & \vb 0.2411 \\
    Cold Start         & 0.7502 & 0.2427 & 0.7502 & 0.2427 & 0.7492 & 0.2431 & 0.7508 & 0.2425 & 0.7566 & 0.2431 \\
    Warmup Start       & 0.7551 & 0.2415 & 0.7556 & 0.2413 & 0.7546 & 0.2417 & 0.7559 & 0.2412 & 0.7529 & 0.2435 \\
    Random Selection   & 0.7563 & 0.2410 & 0.7570 & 0.2411 & 0.7553 & 0.2413 & 0.7565 & 0.2410 & 0.7576 & 0.2408 \\
    K-Means Clustering & 0.7553 & 0.2414 & 0.7558 & 0.2413 & 0.7555 & 0.2415 & 0.7564 & 0.2411 & 0.7550 & 0.2481 \\
    EL2N Selection     & 0.7559 & 0.2412 & 0.7574 & 0.2410 & 0.7569 & 0.2411 & 0.7576 & 0.2407 & 0.7600 & 0.2479 \\
    \best{Ours (DIET)} & \ob \textbf{0.7572} & \ob 0.2409 & \ob \underline{0.7581} & \ob 0.2410 & \ob \underline{0.7572} & \ob 0.2410 & \ob \underline{0.7584} & \ob 0.2406 & \ob \underline{0.7617} & \ob 0.2436 \\

    \specialrule{0.5pt}{1pt}{0pt}

    \multicolumn{11}{c}{\textit{Taobao Dataset} (1.68\%)} \\ 
    \specialrule{0.5pt}{0pt}{0pt}
    \best{Full Data}   & \vb \textbf{0.7146} & \vb 0.1750 & \vb \textbf{0.7166} & \vb 0.1747 & \vb \textbf{0.7113} & \vb 0.1755 & \vb \underline{0.7070} & \vb 0.1766 & \vb \textbf{0.7165} & \vb 0.1766 \\
    Cold Start         & 0.6825 & 0.1786 & 0.6827 & 0.1786 & 0.6754 & 0.1791 & 0.6862 & 0.1787 & 0.6917 & 0.1787 \\
    Warmup Start       & 0.6939 & 0.1776 & 0.6941 & 0.1774 & 0.6972 & 0.1775 & 0.7021 & 0.1767 & 0.7113 & 0.1783 \\
    Random Selection   & 0.6985 & 0.1771 & 0.6995 & 0.1770 & 0.7060 & 0.1761 & 0.7064 & 0.1762 & 0.7060 & 0.1823 \\
    K-Means Clustering & 0.6997 & 0.1772 & 0.7005 & 0.1769 & 0.6978 & 0.1778 & 0.7033 & 0.1766 & 0.7109 & 0.1816 \\
    EL2N Selection     & 0.6977 & 0.1771 & 0.7000 & 0.1769 & 0.7061 & 0.1762 & 0.7082 & 0.1760 & 0.7078 & 0.1805 \\
    \best{Ours (DIET)} & \ob \underline{0.7119} & \ob 0.1754 & \ob \underline{0.7126} & \ob 0.1755 & \ob \underline{0.7100} & \ob 0.1757 & \ob \textbf{0.7101} & \ob 0.1759 & \ob \underline{0.7142} & \ob 0.1793 \\
    
    \specialrule{1.2pt}{1pt}{0pt}
    \end{tabularx}
\end{table*}

\vspace{-0.5\baselineskip}

%% file: table/ablation.tex
\begin{table}[h!]
    \centering
    \vspace{0.5\baselineskip}
    \small
    \caption{\textbf{Ablation Study.} Impact of Memory Fusion (MF), Target Selection (TS), and Memory Selection (MS).}
    \label{tab:ablation_optimized}
    
    \newcommand{\ok}{\textcolor{green!60!black}{\ding{51}}}
    \newcommand{\ko}{\textcolor{red!80}{\ding{55}}}
    \definecolor{OursBlue}{HTML}{EEF3FA}
    \newcommand{\ob}{\cellcolor{OursBlue}}
    
    \renewcommand{\arraystretch}{1.2}
    \setlength\tabcolsep{2pt} 
    
    \begin{tabularx}{\columnwidth}{lcc | *{6}{>{\centering\arraybackslash}X}}
        \specialrule{1.2pt}{0pt}{1pt} 
        
        \multicolumn{3}{c|}{\textbf{Variants}} & \multicolumn{2}{c}{\textbf{KuaiRand}} & \multicolumn{2}{c}{\textbf{Tmall}} & \multicolumn{2}{c}{\textbf{Taobao}} \\
        \cmidrule(lr){1-3} \cmidrule(lr){4-5} \cmidrule(lr){6-7} \cmidrule(lr){8-9}
        
        MF & TS & MS & AUC$\uparrow$ & LogLoss$\downarrow$ & AUC$\uparrow$ & LogLoss$\downarrow$ & AUC$\uparrow$ & LogLoss$\downarrow$ \\
        \specialrule{0.5pt}{1pt}{1pt} 

        \ko & \ok & \ok & 0.7948 & 0.3825 & 0.7598 & 0.2470 & 0.7108 & 0.1808 \\
        \ok & \ko & \ok & 0.7949 & 0.3816 & 0.7611 & 0.2421 & 0.7137 & 0.1797 \\
        \ok & \ok & \ko & 0.7930 & 0.3823 & 0.7598 & 0.2409 & 0.7127 & 0.1807 \\
        \ko & \ko & \ko & 0.7947 & 0.3818 & 0.7590 & 0.2414 & 0.7127 & 0.1793 \\
        
        \addlinespace[2pt]
        
        \ob \textbf{\ok} & \ob \textbf{\ok} & \ob \textbf{\ok} & \ob \textbf{0.7959} & \ob \textbf{0.3817} & \ob \textbf{0.7617} & \ob \textbf{0.2436} & \ob \textbf{0.7142} & \ob \textbf{0.1793} \\
        \specialrule{1.2pt}{1pt}{0pt} 
    \end{tabularx}
\end{table}

%% file: table/complexity.tex
\begin{table}[t!]
    \centering
    \small
    \caption{\textbf{Empirical Time and Memory Overhead.} Comparison of training time and peak GPU memory usage across three stages. Measurement were conducted on the same GPU.}
    \label{tab:empirical_complexity}

    \vspace{-5pt}
    
    \definecolor{VictimBrown}{HTML}{EFEBE7}
    \definecolor{OursBlue}{HTML}{EEF3FA}
    \newcommand{\vb}{\cellcolor{VictimBrown}}
    \newcommand{\ob}{\cellcolor{OursBlue}}
    
    \renewcommand{\arraystretch}{1.2}
    \setlength\tabcolsep{4pt}

    \begin{tabularx}{\columnwidth}{l | cc | cc | cc}
        \specialrule{1.2pt}{0pt}{1pt}
        \textbf{Dataset} & \multicolumn{2}{c|}{\textbf{Full Training}} & \multicolumn{2}{c|}{\textbf{DIET Distill.}} & \multicolumn{2}{c}{\textbf{Distilled Train.}} \\
        \cmidrule(lr){2-3} \cmidrule(lr){4-5} \cmidrule(lr){6-7}
        & Time/ep & Mem. & Time/ep & Mem. & Time/ep & Mem. \\
        \specialrule{0.5pt}{1pt}{1pt}

        KuaiRand & \vb 10.4m & \vb 4.5GB & 0.4m & 24.8GB & \ob 0.1m & \ob 4.5GB \\
        Tmall    & \vb 7.0m & \vb 9.9GB & 1.1m & 32.2GB & \ob 0.1m & \ob 9.9GB \\
        Taobao   & \vb 5.3m & \vb 16.1GB & 10.0m & 51.9GB & \ob 0.2m & \ob 16.1GB \\

        \specialrule{1.2pt}{1pt}{0pt}
        \multicolumn{7}{l}{\footnotesize \textit{Note: Time/ep denotes minutes per epoch. Mem. denotes peak GPU memory (GB).}} \\
    \end{tabularx}
\end{table}

%% file: content/conclusion.tex
\vspace{4pt}
\section{CONCLUSION}
\vspace{4pt}

In this paper, we formalize data-efficient model development in recommender systems under continual training as a \emph{streaming dataset distillation} problem, where a compact dataset is required to approximate the training behavior induced by full historical data streams.
We propose \textbf{DIET}, a framework that constructs and continually refines a compact synthetic dataset as an evolving boundary memory through task-conditioned initialization, continual memory fusion, and influence-guided optimization.
By explicitly modeling training-influential interactions and selectively updating synthetic memory over time, DIET preserves long-term training dynamics that are typically lost under static data reduction.
Extensive experiments on differenct datasets demonstrate that DIET closely approximates full-data training using only a small fraction of historical interactions, while generalizing robustly across heterogeneous model architectures and distillation ratios.
These results indicate that streaming dataset distillation can serve as a practical foundation for efficient and reliable model iteration in large-scale recommender systems.

%% file: content/appendix.tex
\newpage
\section{APPENDIX}

\subsection{DATASET TEMPORAL PARTITIONS}
\label{app:data_splits}

This appendix provides the detailed chronological partitions used in our continual learning protocol.

\paragraph{Tmall.}
The Tmall dataset spans a total of 125 days of user interactions. 
We divide it into three historical stages covering 5/01–5/31, 6/01–6/30, and 7/01–7/31, respectively. 
These are followed by a subsequent stage spanning 8/01–8/31, which represents future interactions in the continual learning setting. 
The remaining two days, 9/01 and 9/02, are used as the validation and test splits, respectively.

\paragraph{Taobao.}
The Taobao dataset covers nine consecutive days of interactions. 
We partition the data into three historical stages, each spanning two days (11/25–11/26, 11/27–11/28, and 11/29–11/30). 
The next day (12/01) serves as the subsequent stage for continual evaluation. 
The final two days, 12/02 and 12/03, are reserved as the validation and test splits.

\paragraph{KuaiRand-1k.}
The KuaiRand-1k dataset spans 23 days of user activity. 
We divide the timeline into two historical stages covering 04/16–04/22 and 04/23–04/29. 
The subsequent stage spans 04/30–05/06 and represents future interactions encountered during continual training. 
The last two days, 05/07 and 05/08, are used as the validation and test splits, respectively.

\subsection{Algorithmic Details of DIET}
\label{app:algorithm}

This appendix provides pseudocode descriptions of DIET for reproducibility.
The algorithms correspond to the two main components introduced in Section~4:
(i) boundary memory initialization, and
(ii) influence-guided optimization via memory addressing.
All symbols follow the definitions in the main text.

\subsubsection{Boundary Memory Initialization}
\label{app:algorithm:init}

Algorithm~\ref{alg:init} summarizes the initialization of the synthetic boundary memory at stage~$t$.
Given the current data block $\mathcal{D}_t$ and teacher checkpoint $\phi_t$, DIET selects task-conditioned training-influential samples using label-conditioned EL2N scoring.
Selected samples are converted into embedding--soft label pairs using the teacher model.
When $t>1$, a subset of historical synthetic data aligned with the current boundary is retained based on influence estimation.
The union forms the initialized synthetic dataset $\mathcal{D}_{syn}^t$.

\subsubsection{Influence-Guided Optimization}
\label{app:algorithm:opt}

Algorithm~\ref{alg:opt} describes the optimization of the synthetic dataset.
At each iteration, DIET performs bidirectional memory addressing to identify hard training targets from the current data block and active synthetic samples.
A proxy model is trained on the active synthetic memory in the inner loop, and the synthetic data are updated in the outer loop using meta-gradients computed on hard targets.
Truncated backpropagation are used for efficient influence estimation.

\input{algorithm/initialization}
\input{algorithm/optimization}

\subsection{ADDITIONAL IMPLEMENTATION DETAILS}
\label{app:details}

This appendix provides additional implementation details for the distillation procedure and memory addressing mechanisms used in DIET, which are omitted from the main text for clarity.

\paragraph{EL2N-based Sample Selection.}
For EL2N-based data selection, samples within each historical data block are ranked according to their EL2N scores computed with respect to the corresponding teacher model checkpoint.
To construct a compressed but informative subset, we select samples around the 90th percentile of the EL2N score distribution, which balances training-influential interactions while avoiding extreme outliers.

\paragraph{Continual Synthetic Memory Construction.}
At each stage $t$, DIET distills a fixed number of new synthetic samples, denoted by $n_t$, from the current data block.
To preserve temporal continuity, we additionally select $n_t$ samples from previously distilled data $\mathcal{D}_{syn}^{1:t-1}$ based on influence estimation.
The newly distilled samples and the selected historical samples are then merged to form the distilled dataset $\mathcal{D}_{syn}^t$ for stage $t$.
This design maintains a balanced contribution from historical and current data while keeping the distilled dataset size controlled.

\paragraph{Bidirectional Memory Addressing.}
During influence-guided memory addressing, the hard target set is selected from the current data block using a fixed ratio of 0.1.
The size of the active synthetic memory is determined as the product of the truncated window size in RaT-BPTT and the number of inner-loop optimization steps.
This setting aligns the scope of synthetic memory updates with the temporal horizon of meta-gradient computation, enabling stable and efficient optimization.

\paragraph{Training Hyperparameters.}
To accommodate the different scales of the datasets while ensuring stable gradient updates, we employ dataset-specific batch sizes. Specifically, the batch sizes are set to 512, 1024, and 4096 for KuaiRand, Tmall, and Taobao, respectively. These configurations are consistently applied to both training on the full datasets and the distilled dataset.

%% file: algorithm/initialization.tex
\begin{algorithm}[t]
\small
\caption{Boundary Memory Initialization at Stage $t$ (Phase~1)}
\label{alg:init}
\KwIn{
Current data block $\mathcal{D}_t$; teacher checkpoint $\phi_t$; historical distilled data $\mathcal{D}_{syn}^{1:t-1}$ (if $t>1$);\\
label combinations set $\mathcal{C}$; selection rule $\textsc{SelectUpperMiddle}(\cdot)$; influence estimation function $\mathcal{A}(\cdot)$.
}
\KwOut{Initialized distilled dataset $\mathcal{D}_{syn}^t$.}

\BlankLine
\textbf{Task-conditioned influential sample selection.}\;
$\mathcal{S}_t \leftarrow \emptyset$\;
\ForEach{$\mathbf{c} \in \mathcal{C}$}{
    $\mathcal{D}_t^{\mathbf{c}} \leftarrow \{(\mathbf{x},\mathbf{y}) \in \mathcal{D}_t \mid \mathbf{y}=\mathbf{c}\}$\;
    \ForEach{$(\mathbf{x},\mathbf{y}) \in \mathcal{D}_t^{\mathbf{c}}$}{
        $s(\mathbf{x},\mathbf{y}) \leftarrow \big\|\sigma(f_{\phi_t}(\mathrm{Embed}(\mathbf{x})))-\mathbf{y}\big\|_2$\tcp*{EL2N}
    }
    $\mathcal{S}_t^{\mathbf{c}} \leftarrow \textsc{SelectUpperMiddle}(\mathcal{D}_t^{\mathbf{c}}, s)$\tcp*{avoid extreme scores}
    $\mathcal{S}_t \leftarrow \mathcal{S}_t \cup \mathcal{S}_t^{\mathbf{c}}$\;
}

\BlankLine
\textbf{Convert to embedding--soft label pairs.}\;
$\mathcal{D}_{cur} \leftarrow \emptyset$\;
\ForEach{$(\mathbf{x},\mathbf{y}) \in \mathcal{S}_t$}{
    $\mathbf{e} \leftarrow \mathrm{Embed}_{\phi_t}(\mathbf{x})$\;
    $\tilde{\mathbf{y}} \leftarrow f_{\phi_t}(\mathrm{Embed}_{\phi_t}(\mathbf{x}))$\tcp*{logits}
    $\mathcal{D}_{cur} \leftarrow \mathcal{D}_{cur} \cup \{(\mathbf{e},\tilde{\mathbf{y}})\}$\;
}

\BlankLine
\textbf{Align and retain historical memory (if applicable).}\;
\eIf{$t = 1$}{
    $\mathcal{D}_{syn}^t \leftarrow \mathcal{D}_{cur}$\;
}{
    \ForEach{$(\mathbf{e}_i,\tilde{\mathbf{y}}_i) \in \mathcal{D}_{syn}^{1:t-1}$}{
        $\alpha_i \leftarrow \mathcal{A}\big((\mathbf{e}_i,\tilde{\mathbf{y}}_i), \mathcal{S}_t\big)$\tcp*{influence score}
    }
    $\widetilde{\mathcal{D}}_{syn}^{1:t-1} \leftarrow \textsc{Top}(\mathcal{D}_{syn}^{1:t-1}, \alpha)$\tcp*{retain aligned memory}
    $\mathcal{D}_{syn}^t \leftarrow \mathcal{D}_{cur} \cup \widetilde{\mathcal{D}}_{syn}^{1:t-1}$\;
}

\Return $\mathcal{D}_{syn}^t$\;
\end{algorithm}

%% file: algorithm/optimization.tex
\begin{algorithm}[h]
\small
\caption{Influence-Guided Optimization at Stage $t$ (Phase~2)}
\label{alg:opt}
\KwIn{
Initialized distilled dataset $\mathcal{D}_{syn}^t$; current data block $\mathcal{D}_t$; teacher checkpoints $\{\phi_{t-1},\phi_t\}$;\\
outer iterations $K$; inner steps $M$; inner LR $\alpha$; meta LR $\beta$;\\
influence score $\mathcal{A}_t(\cdot,\cdot)$; ghost-clipping operator $\textsc{GhostDot}(\cdot,\cdot)$; RaT-BPTT truncation policy.
}
\KwOut{Refined distilled dataset $\mathcal{D}_{syn}^t$.}

\BlankLine
\For{$k = 1$ \KwTo $K$}{
    \tcp{(1) Bidirectional memory addressing}
    Sample a minibatch $\mathcal{B}_t \subset \mathcal{D}_t$\;
    Let $\mathcal{M}_t \leftarrow \mathcal{D}_{syn}^t$\tcp*{current synthetic memory}

    \BlankLine
    \tcp{Hard target selection: $\mathcal{B}_t^{hard}$}
    \ForEach{$z \in \mathcal{B}_t$}{
        $\text{Deficiency}(z) \leftarrow \sum\limits_{x \in \mathcal{M}_t} \mathcal{A}_t(x,z)$\;
    }
    $\mathcal{B}_t^{hard} \leftarrow \textsc{Bottom}(\mathcal{B}_t,\text{Deficiency})$\tcp*{smallest Deficiency}

    \BlankLine
    \tcp{Active memory selection: $\mathcal{M}_t^{active}$}
    \ForEach{$x \in \mathcal{M}_t$}{
        $\text{Resp}(x) \leftarrow \sum\limits_{z \in \mathcal{B}_t^{hard}} \mathcal{A}_t(x,z)$\;
    }
    $\mathcal{M}_t^{active} \leftarrow \textsc{Top}(\mathcal{M}_t,\text{Resp})$\tcp*{largest Resp}

    \BlankLine
    \tcp{(2) Bi-level update with RaT-BPTT}
    \eIf{$t = 1$}{
        Initialize proxy parameters $\theta_0$ randomly\;
    }{
        Initialize proxy parameters $\theta_0 \leftarrow \phi_{t-1}$\tcp*{stage-aware init}
    }

    \BlankLine
    \tcp{Inner loop: train proxy on active memory}
    \For{$m = 1$ \KwTo $M$}{
        $\theta_m \leftarrow \theta_{m-1} - \alpha \nabla_{\theta}\mathcal{L}_{syn}(\theta_{m-1}; \mathcal{M}_t^{active})$\;
    }

    \BlankLine
    \tcp{Outer loop: update distilled data}
    $\mathcal{L}_{meta} \leftarrow \mathcal{L}_{real}(\theta_M; \mathcal{B}_t^{hard})$\;
    Update $\mathcal{D}_{syn}^t \leftarrow \mathcal{D}_{syn}^t - \beta \cdot \textsc{RaT\mbox{-}BPTTGrad}(\mathcal{L}_{meta}, \mathcal{D}_{syn}^t)$\tcp*{random truncated backpropagation through time}

}
\Return $\mathcal{D}_{syn}^t$\;
\end{algorithm}

%% file: reference.bib
@inproceedings{generative-recommender-survey1-10.1145/3701716.3715865,
author = {Wang, Hao and Guo, Wei and Zhang, Luankang and Chin, Jin Yao and Ye, Yufei and Guo, Huifeng and Liu, Yong and Lian, Defu and Tang, Ruiming and Chen, Enhong},
title = {Generative Large Recommendation Models: Emerging Trends in LLMs for Recommendation},
year = {2025},
isbn = {9798400713316},
publisher = {Association for Computing Machinery},
address = {New York, NY, USA},
url = {https://doi.org/10.1145/3701716.3715865},
doi = {10.1145/3701716.3715865},
booktitle = {Companion Proceedings of the ACM on Web Conference 2025},
pages = {49–52},
numpages = {4},
location = {Sydney NSW, Australia},
series = {WWW '25}
}

@article{generative-recommender-survey2-202512.0203,
	doi = {10.20944/preprints202512.0203.v1},
	url = {https://doi.org/10.20944/preprints202512.0203.v1},
	year = 2025,
	month = {December},
	publisher = {Preprints},
	author = {Xiaopeng Li and Bo Chen and Junda She and Shiteng Cao and You Wang and Qinlin Jia and Haiying He and Zheli Zhou and Zhao Liu and Ji Liu and Zhiyang Zhang and Yu Zhou and Guoping Tang and Yiqing Yang and Chengcheng Guo and Si Dong and Kuo Cai and Pengyue Jia and Maolin Wang and Wanyu Wang and Shiyao Wang and Xinchen Luo and Qigen Hu and Qiang Luo and Xiao Lv and Chaoyi Ma and Ruiming Tang and Kun Gai and Guorui Zhou and Xiangyu Zhao},
	title = {A Survey of Generative Recommendation from a Tri-Decoupled Perspective: Tokenization, Architecture, and Optimization},
	journal = {Preprints}
}

@article{GCRank-ni2025gcrank,
  title={GCRank: A Generative Contextual Comprehension Paradigm for Takeout Ranking Model},
  author={Ni, Ziheng and Liu, Congcong and Shang, Cai and Sun, Yiming and Li, Junjie and Fang, Zhiwei and Chen, Guangpeng and Li, Jian and Zhang, Zehua and Peng, Changping and others},
  journal={arXiv preprint arXiv:2601.02361},
  year={2025}
}

@inproceedings{FM-rendle2010factorization,
  title={Factorization machines},
  author={Rendle, Steffen},
  booktitle={2010 IEEE International conference on data mining},
  pages={995--1000},
  year={2010},
  organization={IEEE}
}

@article{DeepFM-guo2017deepfm,
  title={DeepFM: a factorization-machine based neural network for CTR prediction},
  author={Guo, Huifeng and Tang, Ruiming and Ye, Yunming and Li, Zhenguo and He, Xiuqiang},
  journal={arXiv preprint arXiv:1703.04247},
  year={2017}
}

@incollection{DCN-wang2017deep,
  title={Deep \& cross network for ad click predictions},
  author={Wang, Ruoxi and Fu, Bin and Fu, Gang and Wang, Mingliang},
  booktitle={Proceedings of the ADKDD'17},
  pages={1--7},
  year={2017}
}

@inproceedings{DCNv2-wang2021dcn,
  title={Dcn v2: Improved deep \& cross network and practical lessons for web-scale learning to rank systems},
  author={Wang, Ruoxi and Shivanna, Rakesh and Cheng, Derek and Jain, Sagar and Lin, Dong and Hong, Lichan and Chi, Ed},
  booktitle={Proceedings of the web conference 2021},
  pages={1785--1797},
  year={2021}
}

@inproceedings{AutoInt-song2019autoint,
  title={Autoint: Automatic feature interaction learning via self-attentive neural networks},
  author={Song, Weiping and Shi, Chence and Xiao, Zhiping and Duan, Zhijian and Xu, Yewen and Zhang, Ming and Tang, Jian},
  booktitle={Proceedings of the 28th ACM international conference on information and knowledge management},
  pages={1161--1170},
  year={2019}
}

@inproceedings{FinalMLP-mao2023finalmlp,
  title={FinalMLP: an enhanced two-stream MLP model for CTR prediction},
  author={Mao, Kelong and Zhu, Jieming and Su, Liangcai and Cai, Guohao and Li, Yuru and Dong, Zhenhua},
  booktitle={Proceedings of the AAAI conference on artificial intelligence},
  volume={37},
  number={4},
  pages={4552--4560},
  year={2023}
}

@inproceedings{DIN-zhou2018deep,
  title={Deep interest network for click-through rate prediction},
  author={Zhou, Guorui and Zhu, Xiaoqiang and Song, Chenru and Fan, Ying and Zhu, Han and Ma, Xiao and Yan, Yanghui and Jin, Junqi and Li, Han and Gai, Kun},
  booktitle={Proceedings of the 24th ACM SIGKDD international conference on knowledge discovery \& data mining},
  pages={1059--1068},
  year={2018}
}

@inproceedings{DIEN-zhou2019deep,
  title={Deep interest evolution network for click-through rate prediction},
  author={Zhou, Guorui and Mou, Na and Fan, Ying and Pi, Qi and Bian, Weijie and Zhou, Chang and Zhu, Xiaoqiang and Gai, Kun},
  booktitle={Proceedings of the AAAI conference on artificial intelligence},
  volume={33},
  number={01},
  pages={5941--5948},
  year={2019}
}

@inproceedings{BST-chen2019behavior,
  title={Behavior sequence transformer for e-commerce recommendation in alibaba},
  author={Chen, Qiwei and Zhao, Huan and Li, Wei and Huang, Pipei and Ou, Wenwu},
  booktitle={Proceedings of the 1st international workshop on deep learning practice for high-dimensional sparse data},
  pages={1--4},
  year={2019}
}

@article{DSIN-feng2019deep,
  title={Deep session interest network for click-through rate prediction},
  author={Feng, Yufei and Lv, Fuyu and Shen, Weichen and Wang, Menghan and Sun, Fei and Zhu, Yu and Yang, Keping},
  journal={arXiv preprint arXiv:1905.06482},
  year={2019}
}

@inproceedings{SIM-pi2020search,
  title={Search-based user interest modeling with lifelong sequential behavior data for click-through rate prediction},
  author={Pi, Qi and Zhou, Guorui and Zhang, Yujing and Wang, Zhe and Ren, Lejian and Fan, Ying and Zhu, Xiaoqiang and Gai, Kun},
  booktitle={Proceedings of the 29th ACM International Conference on Information \& Knowledge Management},
  pages={2685--2692},
  year={2020}
}

@inproceedings{DV365-lyu2025dv365,
  title={DV365: Extremely Long User History Modeling at Instagram},
  author={Lyu, Wenhan and Tyagi, Devashish and Yang, Yihang and Li, Ziwei and Somani, Ajay and Shanmugasundaram, Karthikeyan and Andrejevic, Nikola and Adeputra, Ferdi and Zeng, Curtis and Singh, Arun K and others},
  booktitle={Proceedings of the 31st ACM SIGKDD Conference on Knowledge Discovery and Data Mining V. 2},
  pages={4717--4727},
  year={2025}
}

@inproceedings{TransActv2-xia2025transact,
  title={TransAct V2: Lifelong User Action Sequence Modeling on Pinterest Recommendation},
  author={Xia, Xue and Joshi, Saurabh and Rajesh, Kousik and Li, Kangnan and Lu, Yangyi and Pancha, Nikil and Badani, Dhruvil and Xu, Jiajing and Eksombatchai, Pong},
  booktitle={Proceedings of the 34th ACM International Conference on Information and Knowledge Management},
  pages={6881--6882},
  year={2025}
}

@article{Monolith-liu2022monolith,
  title={Monolith: real time recommendation system with collisionless embedding table},
  author={Liu, Zhuoran and Zou, Leqi and Zou, Xuan and Wang, Caihua and Zhang, Biao and Tang, Da and Zhu, Bolin and Zhu, Yijie and Wu, Peng and Wang, Ke and others},
  journal={arXiv preprint arXiv:2209.07663},
  year={2022}
}

@article{Foundation-Expert-Meta-li2025realizing,
  title={Realizing Scaling Laws in Recommender Systems: A Foundation-Expert Paradigm for Hyperscale Model Deployment},
  author={Li, Dai and Course, Kevin and Li, Wei and Li, Hongwei and Hua, Jie and Chen, Yiqi and Zhu, Zhao and Jian, Rui and Cao, Xuan and Xue, Bi and others},
  journal={arXiv preprint arXiv:2508.02929},
  year={2025}
}

@article{LUM-yan2025unlocking,
  title={Unlocking Scaling Law in Industrial Recommendation Systems with a Three-step Paradigm based Large User Model},
  author={Yan, Bencheng and Liu, Shilei and Zeng, Zhiyuan and Wang, Zihao and Zhang, Yizhen and Yuan, Yujin and Liu, Langming and Liu, Jiaqi and Wang, Di and Su, Wenbo and others},
  journal={arXiv preprint arXiv:2502.08309},
  year={2025}
}

@article{OneRec-zhou2025onerec,
  title={OneRec Technical Report},
  author={Zhou, Guorui and Deng, Jiaxin and Zhang, Jinghao and Cai, Kuo and Ren, Lejian and Luo, Qiang and Wang, Qianqian and Hu, Qigen and Huang, Rui and Wang, Shiyao and others},
  journal={arXiv preprint arXiv:2506.13695},
  year={2025}
}

@article{WuKong-zhang2024wukong,
  title={Wukong: Towards a scaling law for large-scale recommendation},
  author={Zhang, Buyun and Luo, Liang and Chen, Yuxin and Nie, Jade and Liu, Xi and Guo, Daifeng and Zhao, Yanli and Li, Shen and Hao, Yuchen and Yao, Yantao and others},
  journal={arXiv preprint arXiv:2403.02545},
  year={2024}
}

@inproceedings{Rankmixer-10.1145/3746252.3761507,
author = {Zhu, Jie and Fan, Zhifang and Zhu, Xiaoxie and Jiang, Yuchen and Wang, Hangyu and Han, Xintian and Ding, Haoran and Wang, Xinmin and Zhao, Wenlin and Gong, Zhen and Yang, Huizhi and Chai, Zheng and Chen, Zhe and Zheng, Yuchao and Chen, Qiwei and Zhang, Feng and Zhou, Xun and Xu, Peng and Yang, Xiao and Wu, Di and Liu, Zuotao},
title = {RankMixer: Scaling Up Ranking Models in Industrial Recommenders},
year = {2025},
isbn = {9798400720406},
publisher = {Association for Computing Machinery},
address = {New York, NY, USA},
url = {https://doi.org/10.1145/3746252.3761507},
doi = {10.1145/3746252.3761507},
booktitle = {Proceedings of the 34th ACM International Conference on Information and Knowledge Management},
pages = {6309–6316},
numpages = {8},
location = {Seoul, Republic of Korea},
series = {CIKM '25}
}

@inproceedings{FuXi-alpha-ye2025fuxi,
  title={FuXi-$\alpha$: Scaling Recommendation Model with Feature Interaction Enhanced Transformer},
  author={Ye, Yufei and Guo, Wei and Chin, Jin Yao and Wang, Hao and Zhu, Hong and Lin, Xi and Ye, Yuyang and Liu, Yong and Tang, Ruiming and Lian, Defu and others},
  booktitle={Companion Proceedings of the ACM on Web Conference 2025},
  pages={557--566},
  year={2025}
}

@article{LLaTTE-xiong2026llatte,
  title={LLaTTE: Scaling Laws for Multi-Stage Sequence Modeling in Large-Scale Ads Recommendation},
  author={Xiong, Lee and Chen, Zhirong and Mayuranath, Rahul and Qiu, Shangran and Ozdemir, Arda and Li, Lu and Hu, Yang and Li, Dave and Ren, Jingtao and Cheng, Howard and others},
  journal={arXiv preprint arXiv:2601.20083},
  year={2026}
}

@inproceedings{HugeCTR-wang2022merlin,
  title={Merlin hugectr: Gpu-accelerated recommender system training and inference},
  author={Wang, Zehuan and Wei, Yingcan and Lee, Minseok and Langer, Matthias and Yu, Fan and Liu, Jie and Liu, Shijie and Abel, Daniel G and Guo, Xu and Dong, Jianbing and others},
  booktitle={Proceedings of the 16th ACM Conference on Recommender Systems},
  pages={534--537},
  year={2022}
}

@inproceedings{Coreset-har2004coresets,
  title={On coresets for k-means and k-median clustering},
  author={Har-Peled, Sariel and Mazumdar, Soham},
  booktitle={Proceedings of the thirty-sixth annual ACM symposium on Theory of computing},
  pages={291--300},
  year={2004}
}

@article{Coreset-continual-learning-lopez2017gradient,
  title={Gradient episodic memory for continual learning},
  author={Lopez-Paz, David and Ranzato, Marc'Aurelio},
  journal={Advances in neural information processing systems},
  volume={30},
  year={2017}
}

@article{Coreset-active-learning-settles2009active,
  title={Active learning literature survey},
  author={Settles, Burr},
  year={2009},
  publisher={University of Wisconsin-Madison Department of Computer Sciences}
}

@article{coreset-continue-learning-aljundi2019gradient,
  title={Gradient based sample selection for online continual learning},
  author={Aljundi, Rahaf and Lin, Min and Goujaud, Baptiste and Bengio, Yoshua},
  journal={Advances in neural information processing systems},
  volume={32},
  year={2019}
}

@article{coreset-yang2022dataset,
  title={Dataset pruning: Reducing training data by examining generalization influence},
  author={Yang, Shuo and Xie, Zeke and Peng, Hanyu and Xu, Min and Sun, Mingming and Li, Ping},
  journal={arXiv preprint arXiv:2205.09329},
  year={2022}
}

@article{dd-survey1-liu2025evolution,
  title={The evolution of dataset distillation: Toward scalable and generalizable solutions},
  author={Liu, Ping and Du, Jiawei},
  journal={arXiv preprint arXiv:2502.05673},
  year={2025}
}

@inproceedings{Delt-shen2025delt,
  title={Delt: A simple diversity-driven earlylate training for dataset distillation},
  author={Shen, Zhiqiang and Sherif, Ammar and Yin, Zeyuan and Shao, Shitong},
  booktitle={Proceedings of the Computer Vision and Pattern Recognition Conference},
  pages={4797--4806},
  year={2025}
}

@article{du2024diversity,
  title={Diversity-driven synthesis: Enhancing dataset distillation through directed weight adjustment},
  author={Du, Jiawei and Hu, Juncheng and Huang, Wenxin and Zhou, Joey Tianyi and others},
  journal={Advances in neural information processing systems},
  volume={37},
  pages={119443--119465},
  year={2024}
}

@article{data-distillation-wang2018dataset,
  title={Dataset distillation},
  author={Wang, Tongzhou and Zhu, Jun-Yan and Torralba, Antonio and Efros, Alexei A},
  journal={arXiv preprint arXiv:1811.10959},
  year={2018}
}

@article{data-distillation-survey1-sachdeva2023data,
  title={Data distillation: A survey},
  author={Sachdeva, Noveen and McAuley, Julian},
  journal={arXiv preprint arXiv:2301.04272},
  year={2023}
}

@article{data-distillation-survey2-geng2023survey,
  title={A survey on dataset distillation: Approaches, applications and future directions},
  author={Geng, Jiahui and Chen, Zongxiong and Wang, Yuandou and Woisetschlaeger, Herbert and Schimmler, Sonja and Mayer, Ruben and Zhao, Zhiming and Rong, Chunming},
  journal={arXiv preprint arXiv:2305.01975},
  year={2023}
}

@article{data-distillation-survey3-liu2025evolution,
  title={The evolution of dataset distillation: Toward scalable and generalizable solutions},
  author={Liu, Ping and Du, Jiawei},
  journal={arXiv preprint arXiv:2502.05673},
  year={2025}
}

@inproceedings{TD3-zhang2025td3,
  title={Td3: Tucker decomposition based dataset distillation method for sequential recommendation},
  author={Zhang, Jiaqing and Yin, Mingjia and Wang, Hao and Li, Yawen and Ye, Yuyang and Lou, Xingyu and Du, Junping and Chen, Enhong},
  booktitle={Proceedings of the ACM on Web Conference 2025},
  pages={3994--4003},
  year={2025}
}

@article{ratbptt-feng2023embarassingly,
  title={Embarassingly simple dataset distillation},
  author={Feng, Yunzhen and Vedantam, Ramakrishna and Kempe, Julia},
  journal={arXiv preprint arXiv:2311.07025},
  year={2023}
}

@article{aat-li2025beyond,
  title={Beyond Random: Automatic Inner-loop Optimization in Dataset Distillation},
  author={Li, Muquan and Gou, Hang and Zhang, Dongyang and Liang, Shuang and Xie, Xiurui and Ouyang, Deqiang and Qin, Ke},
  journal={arXiv preprint arXiv:2510.04838},
  year={2025}
}

@article{DC-zhao2020dataset,
  title={Dataset condensation with gradient matching},
  author={Zhao, Bo and Mopuri, Konda Reddy and Bilen, Hakan},
  journal={arXiv preprint arXiv:2006.05929},
  year={2020}
}

@inproceedings{DSA-zhao2021dataset,
  title={Dataset condensation with differentiable siamese augmentation},
  author={Zhao, Bo and Bilen, Hakan},
  booktitle={International Conference on Machine Learning},
  pages={12674--12685},
  year={2021},
  organization={PMLR}
}

@inproceedings{DCctr-wang2023gradient,
  title={Gradient matching for categorical data distillation in ctr prediction},
  author={Wang, Cheng and Sun, Jiacheng and Dong, Zhenhua and Li, Ruixuan and Zhang, Rui},
  booktitle={Proceedings of the 17th ACM Conference on Recommender Systems},
  pages={161--170},
  year={2023}
}

@inproceedings{MTT1-cazenavette2022dataset,
  title={Dataset distillation by matching training trajectories},
  author={Cazenavette, George and Wang, Tongzhou and Torralba, Antonio and Efros, Alexei A and Zhu, Jun-Yan},
  booktitle={Proceedings of the IEEE/CVF Conference on Computer Vision and Pattern Recognition},
  pages={4750--4759},
  year={2022}
}

@inproceedings{MTT2-du2023minimizing,
  title={Minimizing the accumulated trajectory error to improve dataset distillation},
  author={Du, Jiawei and Jiang, Yidi and Tan, Vincent YF and Zhou, Joey Tianyi and Li, Haizhou},
  booktitle={Proceedings of the IEEE/CVF conference on computer vision and pattern recognition},
  pages={3749--3758},
  year={2023}
}

@inproceedings{MTT3-liu2024dataset,
  title={Dataset distillation by automatic training trajectories},
  author={Liu, Dai and Gu, Jindong and Cao, Hu and Trinitis, Carsten and Schulz, Martin},
  booktitle={European Conference on Computer Vision},
  pages={334--351},
  year={2024},
  organization={Springer}
}

@inproceedings{MTT4-zhong2025towards,
  title={Towards stable and storage-efficient dataset distillation: Matching convexified trajectory},
  author={Zhong, Wenliang and Tang, Haoyu and Zheng, Qinghai and Xu, Mingzhu and Hu, Yupeng and Guan, Weili},
  booktitle={Proceedings of the Computer Vision and Pattern Recognition Conference},
  pages={25581--25589},
  year={2025}
}

@inproceedings{DBLP:conf/iclr/KaiserSRK25,
  author       = {Johannes Kaiser and
                  Kristian Schwethelm and
                  Daniel Rueckert and
                  Georgios Kaissis},
  title        = {Laplace Sample Information: Data Informativeness Through a Bayesian
                  Lens},
  booktitle    = {The Thirteenth International Conference on Learning Representations,
                  {ICLR} 2025, Singapore, April 24-28, 2025},
  publisher    = {OpenReview.net},
  year         = {2025},
  url          = {https://openreview.net/forum?id=qO6dk9KfIp},
  bibsource    = {dblp computer science bibliography, https://dblp.org}
}

@inproceedings{DBLP:conf/cvpr/ShenS0S25,
  author       = {Zhiqiang Shen and
                  Ammar Sherif and
                  Zeyuan Yin and
                  Shitong Shao},
  title        = {{DELT:} {A} Simple Diversity-driven EarlyLate Training for Dataset
                  Distillation},
  booktitle    = {{IEEE/CVF} Conference on Computer Vision and Pattern Recognition,
                  {CVPR} 2025, Nashville, TN, USA, June 11-15, 2025},
  pages        = {4797--4806},
  publisher    = {Computer Vision Foundation / {IEEE}},
  year         = {2025},
  url          = {https://openaccess.thecvf.com/content/CVPR2025/html/Shen\_DELT\_A\_Simple\_Diversity-driven\_EarlyLate\_Training\_for\_Dataset\_Distillation\_CVPR\_2025\_paper.html},
  doi          = {10.1109/CVPR52734.2025.00452},
  bibsource    = {dblp computer science bibliography, https://dblp.org}
}

@inproceedings{EL2N-DBLP:conf/nips/PaulGD21,
  author       = {Mansheej Paul and
                  Surya Ganguli and
                  Gintare Karolina Dziugaite},
  editor       = {Marc'Aurelio Ranzato and
                  Alina Beygelzimer and
                  Yann N. Dauphin and
                  Percy Liang and
                  Jennifer Wortman Vaughan},
  title        = {Deep Learning on a Data Diet: Finding Important Examples Early in
                  Training},
  booktitle    = {Advances in Neural Information Processing Systems 34: Annual Conference
                  on Neural Information Processing Systems 2021, NeurIPS 2021, December
                  6-14, 2021, virtual},
  pages        = {20596--20607},
  year         = {2021},
  url          = {https://proceedings.neurips.cc/paper/2021/hash/ac56f8fe9eea3e4a365f29f0f1957c55-Abstract.html},
  bibsource    = {dblp computer science bibliography, https://dblp.org}
}

@inproceedings{TracIN-DBLP:conf/nips/PruthiLKS20,
  author       = {Garima Pruthi and
                  Frederick Liu and
                  Satyen Kale and
                  Mukund Sundararajan},
  editor       = {Hugo Larochelle and
                  Marc'Aurelio Ranzato and
                  Raia Hadsell and
                  Maria{-}Florina Balcan and
                  Hsuan{-}Tien Lin},
  title        = {Estimating Training Data Influence by Tracing Gradient Descent},
  booktitle    = {Advances in Neural Information Processing Systems 33: Annual Conference
                  on Neural Information Processing Systems 2020, NeurIPS 2020, December
                  6-12, 2020, virtual},
  year         = {2020},
  url          = {https://proceedings.neurips.cc/paper/2020/hash/e6385d39ec9394f2f3a354d9d2b88eec-Abstract.html},
  bibsource    = {dblp computer science bibliography, https://dblp.org}
}

@inproceedings{Shapley-DBLP:conf/iclr/WangMS025,
  author       = {Jiachen T. Wang and
                  Prateek Mittal and
                  Dawn Song and
                  Ruoxi Jia},
  title        = {Data Shapley in One Training Run},
  booktitle    = {The Thirteenth International Conference on Learning Representations,
                  {ICLR} 2025, Singapore, April 24-28, 2025},
  publisher    = {OpenReview.net},
  year         = {2025},
  url          = {https://openreview.net/forum?id=HD6bWcj87Y},
  bibsource    = {dblp computer science bibliography, https://dblp.org}
}

@inproceedings{FinalNet-DBLP:conf/sigir/ZhuJCDLDTZ23,
  author       = {Jieming Zhu and
                  Qinglin Jia and
                  Guohao Cai and
                  Quanyu Dai and
                  Jingjie Li and
                  Zhenhua Dong and
                  Ruiming Tang and
                  Rui Zhang},
  editor       = {Hsin{-}Hsi Chen and
                  Wei{-}Jou (Edward) Duh and
                  Hen{-}Hsen Huang and
                  Makoto P. Kato and
                  Josiane Mothe and
                  Barbara Poblete},
  title        = {{FINAL:} Factorized Interaction Layer for {CTR} Prediction},
  booktitle    = {Proceedings of the 46th International {ACM} {SIGIR} Conference on
                  Research and Development in Information Retrieval, {SIGIR} 2023, Taipei,
                  Taiwan, July 23-27, 2023},
  pages        = {2006--2010},
  publisher    = {{ACM}},
  year         = {2023},
  url          = {https://doi.org/10.1145/3539618.3591988},
  doi          = {10.1145/3539618.3591988},
  bibsource    = {dblp computer science bibliography, https://dblp.org}
}

@inproceedings{FuxiCTR-zhu2021open,
  title={Open benchmarking for click-through rate prediction},
  author={Zhu, Jieming and Liu, Jinyang and Yang, Shuai and Zhang, Qi and He, Xiuqiang},
  booktitle={Proceedings of the 30th ACM international conference on information \& knowledge management},
  pages={2759--2769},
  year={2021}
}

@article{difficulty-aligned-dd1-guo2023towards,
  title={Towards lossless dataset distillation via difficulty-aligned trajectory matching},
  author={Guo, Ziyao and Wang, Kai and Cazenavette, George and Li, Hui and Zhang, Kaipeng and You, Yang},
  journal={arXiv preprint arXiv:2310.05773},
  year={2023}
}

@article{difficulty-aligned-dd2-lee2024selmatch,
  title={Selmatch: Effectively scaling up dataset distillation via selection-based initialization and partial updates by trajectory matching},
  author={Lee, Yongmin and Chung, Hye Won},
  journal={arXiv preprint arXiv:2406.18561},
  year={2024}
}

@article{difficulty-aligned-dd3-wang2024not,
  title={Not all samples should be utilized equally: Towards understanding and improving dataset distillation},
  author={Wang, Shaobo and Yang, Yantai and Wang, Qilong and Li, Kaixin and Zhang, Linfeng and Yan, Junchi},
  journal={arXiv preprint arXiv:2408.12483},
  year={2024}
}

@inproceedings{00-wang-shenp,
  title={P-Law: Predicting Quantitative Scaling Law with Entropy Guidance in Large Recommendation Models},
  author={Shen, Tingjia and Wang, Hao and Wu, Chuhan and Chin, Jin Yao and Guo, Wei and Liu, Yong and Guo, Huifeng and Lian, Defu and Tang, Ruiming and Chen, Enhong},
  booktitle={The Thirty-ninth Annual Conference on Neural Information Processing Systems}
}

@article{02-wang-xie2024breaking,
  title={Breaking Determinism: Fuzzy Modeling of Sequential Recommendation Using Discrete State Space Diffusion Model},
  author={Xie, Wenjia and Wang, Hao and Zhang, Luankang and Zhou, Rui and Lian, Defu and Chen, Enhong},
  journal={arXiv preprint arXiv:2410.23994},
  year={2024}
}

@article{03-wang-00-yin2025feature,
  title={From Feature Interaction to Feature Generation: A Generative Paradigm of CTR Prediction Models},
  author={Yin, Mingjia and Pan, Junwei and Wang, Hao and Wang, Ximei and Zhang, Shangyu and Jiang, Jie and Lian, Defu and Chen, Enhong},
  journal={arXiv preprint arXiv:2512.14041},
  year={2025}
}

@inproceedings{04-wang-01-xie2025breaking,
  title={Breaking the Bottleneck: User-Specific Optimization and Real-Time Inference Integration for Sequential Recommendation},
  author={Xie, Wenjia and Wang, Hao and Fang, Minghao and Yu, Ruize and Guo, Wei and Liu, Yong and Lian, Defu and Chen, Enhong},
  booktitle={Proceedings of the 31st ACM SIGKDD Conference on Knowledge Discovery and Data Mining V. 2},
  pages={3333--3343},
  year={2025}
}

@inproceedings{05-wang-02-xu2025multi,
  title={Multi-granularity interest retrieval and refinement network for long-term user behavior modeling in ctr prediction},
  author={Xu, Xiang and Wang, Hao and Guo, Wei and Zhang, Luankang and Yang, Wanshan and Yu, Runlong and Liu, Yong and Lian, Defu and Chen, Enhong},
  booktitle={Proceedings of the 31st ACM SIGKDD Conference on Knowledge Discovery and Data Mining V. 1},
  pages={2745--2755},
  year={2025}
}

@inproceedings{07-wang-05-zhou2025multi,
  title={MIT: A Multi-Tower Information Transfer Framework Based on Hierarchical Task Relationship Modeling},
  author={Zhou, Rui and Wang, Hao and Guo, Wei and Jia, Qinglin and Xie, Wenjia and Xu, Xiang and Liu, Yong and Lian, Defu and Chen, Enhong},
  booktitle={Companion Proceedings of the ACM on Web Conference 2025},
  pages={651--660},
  year={2025}
}

@inproceedings{08-wang-06-wang2025universal,
  title={A universal framework for compressing embeddings in ctr prediction},
  author={Wang, Kefan and Wang, Hao and Song, Kenan and Guo, Wei and Cheng, Kai and Li, Zhi and Liu, Yong and Lian, Defu and Chen, Enhong},
  booktitle={International Conference on Database Systems for Advanced Applications},
  pages={84--100},
  year={2025},
  organization={Springer}
}
